\documentclass[letterpaper]{aa540}
\usepackage{graphicx1}
\usepackage{txfonts}
\usepackage{natbib}
\usepackage{aalongtable}
\usepackage{subeqn}
\bibpunct{(}{)}{;}{a}{}{,}

\def\deg{\hbox{$^\circ$}}

\def\ga{\mathrel{\hbox{\rlap{\hbox{\lower4pt\hbox{$\sim$}}}\hbox{$>$}}}}
\def\la{\mathrel{\hbox{\rlap{\hbox{\lower4pt\hbox{$\sim$}}}\hbox{$<$}}}}
\def\msunyr{$M$ \mbox{$_{\normalsize\odot}$}\rm{yr}^{-1}}
\def\msun{$M$ \mbox{$_{\normalsize\odot}$}}
\def\lsun{$L$ \mbox{$_{\normalsize\odot}$}}
\def\kms{\,km~s$^{-1}$}
\begin{document}
   \title{Asphericity and clumpiness in the winds of Luminous Blue Variables}

   \subtitle{}

   \author{Ben Davies
          \inst{1}
          \and
          Ren\'{e} D. Oudmaijer\inst{1}
	    \and
	    Jorick S. Vink\inst{2}
          }

   \offprints{B. Davies}

   \institute{School of Physics \& Astronomy, University of Leeds,
              Woodhouse Lane, Leeds LS2 9JT, UK\\
              \email{bd@ast.leeds.ac.uk; roud@ast.leeds.ac.uk}
         \and
             Blackett Laboratory, Imperial College, Prince Consort Road, London SW7 2BZ, UK \\
		  \email{j.vink@imperial.ac.uk}
             }

   \date{Received 31 January 2005 / Accepted 8 May 2005}

   \abstract{We present the first systematic spectropolarimetric study
of Luminous Blue Variables (LBVs) in the Galaxy and the Magellanic
Clouds, in order to investigate the geometries of their winds. We find
that at least half of our sample show changes in polarization across
the strong H$\alpha$ emission line, indicating that the light from the
stars is intrinsically polarized and therefore that asphericity
already exists at the base of the wind. Multi-epoch spectropolarimetry
on four targets reveals variability in their intrinsic
polarization. Three of these, AG Car, HR Car and P Cyg, show a
position angle (PA) of polarization which appears random with
time. Such behaviour can be explained by the presence of strong
wind-inhomogeneities, or `clumps' within the wind. Only one star, R
127, shows variability at a constant PA, and hence evidence for
axi-symmetry as well as clumpiness.  However, if viewed at low
inclination, and at limited temporal sampling, such a wind would
produce a seemingly random polarization of the type observed in the
other three stars. Time-resolved spectropolarimetric monitoring of
LBVs is therefore required to determine if LBV winds are axi-symmetric
in general.

\hspace{5mm} The high fraction of LBVs ($>$ 50\%) showing intrinsic
polarization is to be compared with the lower $\sim$ 20-25 \% for
similar studies of their evolutionary neighbours, O supergiants and
Wolf-Rayet stars. We anticipate that this higher incidence is due to
the lower effective gravities of the LBVs, coupled with their variable
temperatures within the bi-stability jump regime. This is also
consistent with the higher incidence of wind asphericity that we find
in LBVs with strong H$\alpha$ emission and recent (last $\sim$ 10
years) strong variability.

   }

   \maketitle
%


\begin {table*}[]
 \begin{center}
  \begin {tabular}{lcccccccc}
	\hline
      	Name 	& \multicolumn{2}{c}{Coords (2000)} & $m_{V}$ & Obs Date & Integration & Slit & S/N & Res. \\ 
		& $\alpha$ & $\delta$		&      & 	&  Time (s)  & Width ('') & & (\AA)  \\
      	\hline \hline
	\emph{Galactic} &	&			&	&	&	&	&	\\
	\object{AG Car} 	& 10 56 11.58 & -60 27 12.8	& 8.1 -- 6.0      & 06.05.94 & 960  & 1.0  &  600 & 1.5  \\
\vspace{2mm}
	'' 	& " 		& "      		& ''	& 10.02.03 & 600  & 1.0  & 500  & 1.1  \\
	\object{HR Car}	& 10 22 53.84 & -59 37 28.4 	& 8.7 -- 7.4      & 16.03.92 &  3200    & 1.0  & 1300     & 1.5     \\
\vspace{2mm}
	''	& " 		& "	      	& ''	& 10.02.03 & 680  & 1.0  & 600  & 1.1  \\
	\object{$\eta$ Car}	& 10 45 03.59 & -59 41 04.3	& 6.2      & 13.02.03 & 12   & 1.0  & 300  & 1.1  \\
	\object{Hen 3-519}	& 10 53 59.66 & -60 26 44.3	& 11.0	& 06.05.94 & 1600   & 1.0  & 500  & 1.5  \\
\vspace{2mm} ''	& '' 		  & ''		& ''      & 10.02.03 & 8880	& 1.0  & 800  & 1.1  \\
	WRA 751	& 11 08 40.4  & -60 42 51 	& 12.3 -- 10.3      &	11.02.03 & 8400 & 1.5  & 300  & 1.3  \\
	\object{P Cyg}	& 20 17 47.2  & +38 01 58.5	      & 4.9 & 10.12.03 & 64     & 1.0     & 700  & 0.8 \\
\vspace{2mm}
	''	& ''	      & ''		      & '' & 13.12.03 & 48     & 1.0     & 600  & 0.8 \\
\vspace{2mm}	\object{HD 160529}	& 17 41 59.03 & -33 30 13.7	& 6.95 -- 6.4  & 11.02.03 & 620	& 1.0  & 450  & 1.1  \\

	\emph{SMC} &		&			&	&	&	&	&	\\
	R 40 	& 01 07 18.22 & -72 28 03.7	& 10.7 -- 9.9      & 14.08.03 & 8400 & 2.5	& 350	& 1.7	\\

\vspace{2mm}	''	& ''	      & ''		     & ''  & 15.08.03 & 4400 & 2.0	& 300	& 1.5	\\
	\emph{LMC} &		&			&	&	&	&	& \\
	\object{S Dor}	&  05 18 14.3 & -69 14 59	     & 11.4 -- 9.0 & 11.02.03 & 4800 & 1.0	& 500	& 1.1 \\
	R 71  & 05 02 07.39 & -71 20 13.1	     & 11.0 -- 9.9 & 12.02.03 & 7200 & 1.5   & 500	& 1.2 \\	
	R 110 & 05 30 51.48 & -69 02 58.7	     & 10.6 -- 9.7 & 12.02.03 & 7200 & 1.5	& 500   & 1.3 \\
	R 116 & 05 31 52.28 & -68 32 38.9	     & 10.6 & 13.02.03 & 7200 & 1.0	& 600	& 1.1 \\
	R 127 & 05 36 43.50 & -69 29 45.0	     & 11.4 -- 8.8 & 13.02.03 & 2880 & 1.0	& 400	& 1.1 \\
\vspace{2mm}
	R 143 & 05 38 51.60 & -69 08 07.1	     & 12.0 -- 10.6 & 14.02.03 & 4800 & 2.0	& 350	& 1.5 \\
	
      \hline
      \end {tabular}
     \end{center}
\caption{Log of the LBV observations. The signal-to-noise (S/N) was
calculated from the noise in flat regions of the continuum. The
spectral resolution was calculated using the FWHM of lines in the arc
calibration spectra taken at the time and position of the observed
star. All observations, including the archive data, were done at the
AAT except \object{P Cyg} which was done at the WHT. All data was
taken using a 1200R grating, except \object{HD 160529} on 18/9/02
which was done using a 600V grating. \label{tab:lbvobs}}

\end {table*}


\begin {table*}[]
 \begin{center}
  \begin {tabular}{lcccccccc}
	\hline
      	Name 	& \multicolumn{2}{c}{Coords (2000)} & Spec & Integration & Slit & S/N & Res. \\ 
		& $\alpha$ & $\delta$		      & Type &	 Time (s)  & Width ('') & & (\AA)  \\
      	\hline \hline
\object{HR 5281}&14 06 25.2&-59 42 57&B0Iae&120&1.5&350&1.2	\\
\object{HR 5027}&13 20 48.3&-55 48 03&B0.5Iae&590&1.5&550&1.2	\\
\object{HR 6142}&16 31 41.8&-41 49 02&B1Iae&70&1.5&600&1.2	\\
\object{HR 4806}&12 38 52.4&-67 11 35&B1Ia&1260&2&600&1.2	\\
\object{HR 4653}&12 14 16.9&-64 24 31&B1.5Iae&530&2&700&1.2	\\
\object{HR 4198}&10 42 40.6&-59 12 57&B2.5Iae&660&1&500&1.1	\\
\object{HR 3940}&09 56 51.7&-54 34 04&B5Ia&100&2&1100&1.2	\\
\object{HR 4611}&12 05 53.6&-65 32 49&B6Iab&600&1.5&450&1.2	\\
\object{HD 183143}&19 27 26.6&+18 17 45&B8Iae&120&1&900&2.1	\\
\object{HD 32034}&04 55 11.1&-67 10 11&B9Iae&960&2&100&1.2	\\
\object{HR 4169}&10 37 27.1&-58 44 00&A0Iae&180&1.5&400&1.2	\\
\object{HR 4228}&10 48 05.4&-59 55 09&A0Ia &200&1&400&1.1	\\
\object{HR 3975}&10 07 20.0&+16 45 46&A0Ia &120&1&800&1.1	\\
\object{HR 4352}&11 48 45.1&-26 44 59&A6Ia&120&1&700&1.1	\\

      \hline
      \end {tabular}
     \end{center}
\caption{Log of the observations of the stars in the spectral
atlas. Columns show the same as Table \ref{tab:lbvobs}, except column
4 which shows the MK spectral type of each of the stars. All
observations were done at the AAT, and all used a 1200R grating with
the exception of HD 183143 which used a 600V grating. All stars were
observed on 15/2/04 with the exception of HD 183143 and HD 32034 which
were observed on 15/9/02 and 15/8/03
respectively. \label{tab:atlasobs}}
\end {table*}

\section{Introduction \label{sec:intro}}

Luminous Blue Variables (LBVs), or S Dor variables, are a class of
evolved massive stars located close to the empirical upper-luminosity
boundary on the H-R diagram, known as the Humphreys-Davidson (HD)
limit \citep[][ hereafter vG01]{H-D94,vG01}. They exhibit photometric
variability on three different scales: microvariations of a few tenths
of a magnitude on timescales of weeks to months; variations of $\sim$
1 mag on timescales of years; and massive eruptions of $\ga$ 2 mag,
observed in only a handful of objects, \object{$\eta$ Car} and
\object{P Cyg} being the most famous
\citep[see][]{Lamers98,Humphreys99}. At visual minimum their spectra
resemble those of early-type supergiants with emission lines of H
\textsc{i}, He \textsc{i} and Fe \textsc{ii}, with temperatures
ranging from 15-30,000K. At visual maximum, their temperatures are
observed to fall to $\sim$ 9,000K, as their radius increases and
atmosphere cools \citep{dK96}. Their mass-loss rates, of $\sim$
$10^{-5} - 10^{-4}~\msunyr$, vary with effective temperature, which
can be understood by radiation pressure on spectral lines
\citep{V-dK02}.  The LBV phase itself is expected to last $\sim
10^{5}$ yrs, and is thought to be a transitional phase experienced by
stars with an initial mass $\ga$ 25$\msun$ prior to the Wolf-Rayet
stage (e.g. Maeder 1997).

With only a handful of exceptions, LBVs are surrounded by expanding
aspherical nebulae \citep{Weis03,Nota95}. These nebulae can be
bipolar, elliptical or irregular; be discrete shells or clumpy, and
are thought to result from the stars' episodic mass-loss
phases. \citet{Nota95} proposed that these nebulae may be formed by a
spherically symmetric wind interacting with a pre-existing density
contrast left over from a prior mass-loss episode, which was
subsequently backed-up with models \citep{Frank95,D-B98,Langer99}.  It
has however also been shown that such nebulae can be formed if the
wind \emph{itself} is axi-symmetric.  A bipolar wind can be created if
the rotation rate is close to break-up and the star's flux is latitude
dependent due to oblateness and gravity-darkening
\citep[e.g.][]{D-O02}. An equatorial wind may exist if the star is
close to the bi-stability jump temperature where the recombination of
Fe \textsc{iv} to Fe \textsc{iii} below the sonic point of the wind
leads to an increased opacity and a marked change in wind properties
\citep{Vink99}. Indeed, such a scenario has been suggested by
\citet{L-P91} and \citet{Pelupessy00} to explain the two-component
wind model of the possibly related B[e] supergiants. However, direct
observational investigations into the present-day wind geometries of
evolved massive stars in general, and LBVs in particular, are thus far
thin on the ground.

Evidence for aspherical stellar winds can be found through
spectropolarimetry. Ionised circumstellar gas in the wind which has a
flattened geometry on the sky, such as in a bi-polar or
equatorially-enhanced flow, (electron) scatters continuum photons
originating from the star. In the optically thin case, the scattered
continuum light is then linearly polarized perpendicular to the plane
of the flow. In the case of an optically thick wind, multiple
scattering effects can mean the light is polarized parallel to the
plane of the flow \citep{Angel69,Wood96}. Emission-line radiation
emanating from the ionized gas, which is formed over a much larger
volume at a larger radius, undergoes less scattering and remains
essentially unpolarized. Therefore a drop in polarization across an
emission line is indicative of aspherical structures within the
line-forming region. As the bulk of the polarization occurs within a
couple of stellar radii where the circumstellar gas is most dense
\citep{Cassinelli87}, this signature tells us about the geometry at
the very base of the wind.

Spectropolarimetric evidence for aspherical stellar winds exists for
the Galactic LBVs \object{AG Car} \citep{Leitherer94,S-L94},
\object{HR Car} \citep{Clampin95}, and \object{P Cyg}
\citep{Taylor91,Nordsieck01}, as well as the Magellanic Cloud LBV
\object{R 127} \citep{S-L93}. These studies deal only with individual
objects, however, and until now no systematic study exists to
determine whether LBVs \emph{in general} undergo aspherical mass loss.
Such studies do exist for O supergiants and Wolf-Rayet stars -- of
which LBVs are possibly the evolutionary mid-point -- but evidence for
significant intrinsic polarization was found in only $\sim$ 25\% and
$\sim$ 20\% of these groups respectively
\citep{Harries02,Harries98}. As LBVs are closer to their modified
Eddington limit, have lower effective gravities, and are in the
bi-stability jump regime, we may expect to find different mass-loss
behaviour. In the first systematic spectropolarimetric study of LBVs,
we will show that they do indeed show a higher proportion of intrinsic
polarization, and hence that their wind geometries in general differ
from those of their evolutionary neighbours.

We begin in Section \ref{sec:obs} by describing the observational
technique and data-reduction steps. We describe in Section
\ref{sec:results} an empirical method for estimating the stars'
temperatures (and hence phase) at the epoch of observation, and
displays the results of the spectropolarimetry. These results are
discussed in Section \ref{sec:disc}.


\section{Observations \& Data Reduction \label{sec:obs}}

The linear spectropolarimetric data were taken during four observing
runs. Three were done at the 3.9m AAT using the 25cm camera of the RGO
Spectrograph in September 2002, February 2003 and August 2003. The
fourth was done at the 4.2m WHT using ISIS in December 2003. The
spectroscopic data was taken on the AAT runs as part of bad-weather
back-up programmes, and exactly the same set-up was used as the
spectropolarimetry data, minus the polarization optics. The weather
was mixed during each run, with some cloud around on some of the
nights data was taken. The seeing varied from sub-arcsecond to $\ga$
2\arcsec ~conditions. Additionally, the AAT archive was searched for
previous spectropolarimetric observations of LBVs. Only high/medium
resolution, previously un-published data was selected.

The instrumental set-up was similar for all the AAT spectropolarimetry
observations. A dekker mask with two holes of size 2.7\arcsec \space
and separation 22.8\arcsec \space was placed in front of the slit in
order to observe the target and the sky simultaneously. A half-wave
plate is used to rotate the polarization of the incoming light, and a
calcite block splits the light into two perpendicularly polarized
beams (the $O$ and the $E$ rays), hence four spectra are recorded --
the $O$ and $E$ beams, of both the target and the sky. One complete
polarization observation consists of one exposure at four different
waveplate positions -- 0\deg \space and 45\deg \space (to measure
Stokes $Q$), 22.5\deg \space and 67.5\deg \space (to measure Stokes
$U$). Several sets of each object were obtained to check for
repeatability of results, and also to avoid saturation of the CCD at
the peak of the H$\alpha$ line. Spectropolarimetric and
zero-polarization standards were observed each night, and calibration
spectra were taken after each target by observing a copper-argon lamp.

The August 2003 and February 2003 runs used a 1200R grating in
conjunction with the \textsc{eev2} CCD windowed to 200$\times$2800,
giving a spectral range of 1250\AA \space centred at 6500\AA. The
September 2002 run used a 600V grating in conjuction with the
\textsc{mitll3} CCD windowed to 200$\times$3584, giving a spectral
range of 2690\AA \space centred at 5600\AA. P Cygni was observed in
December 2003 at the WHT, and the set-up is described in
\citet{Vink03}. The {\sc marconi2} CCD was used in conjunction with
the 1200R grating, which gave a range of 1055\AA \space centred on
6500\AA. A log of the observations, including integration times and
spectral resolution achieved is shown in tables \ref{tab:lbvobs} and
\ref{tab:atlasobs}.

The reduction steps included bias subtraction, flat-fielding,
cosmic-ray removal, chip linearity correction, spectrum extraction and
wavelength calibration. The sky spectra were subtracted from the
object spectra in order to compensate for atmospheric features and
extended nebulous emission which may contaminate the star aperture.
These steps were done using the \textsc{figaro} package maintained by
Starlink. Determination of the Stokes parameters was done using the
Time-Series/Polarimetry (\textsc{tsp}) package, also maintained by
Starlink. From the Stokes parameters, the degree of polarization $P$
and position-angle (PA) $\theta$ were found from,

\begin{equation}
P = \sqrt{ Q^{2} + U^{2} }
\label{equ:p}
\end{equation}

\begin{equation}
\theta = \frac{1}{2} \arctan{ \frac{U}{Q} }
\label{equ:th}
\end{equation}

PA calibration and wave-plate chromicity were corrected for using the
Starlink package \textsc{polmap}, by fitting a 4th-degree polynomial
to the spectra of polarized standards. The high-frequency `ripple'
observed in high S/N spectropolarimetry data, caused by internal
reflections within the waveplate assembly \citep{A-H01,H-H96}, was
corrected for using a Fourier transform algorithm, also within
\textsc{polmap}. Instrumental polarization, determined from the
zero-polarization standards, was found to be less than 0.1\%, with the
origin in $Q-U$ space within the 1$\sigma$ rms scatter. As we are only
interested in the internal precision of the instrument in order to
detect changes in polarization across spectral features, this has not
been corrected for.

The residuals between the measured Stokes parameters of polarized
standards and their literature values showed a 1$\sigma$ scatter
around zero of $\sim$ 0.2\%. The uncertainty in position-angle $\Delta
\theta \sim 29 \deg \times \Delta P / P$, and is around 1.5\deg \space
for $P=4\%$ and around 6\deg \space for $P=1\%$.

There were several extra steps required in the reduction. The $O$ and
$E$ spectra, as well as spectra from different waveplate positions,
were found to be spectrally misaligned with each other by a few tenths
of a pixel. As the determination of the Stokes parameters essentially
requires the spectra be divided by one another, this introduces
artifacts in the polarization spectrum around sharp features. To
compensate for this the spectra were cross-correlated with each other
and realigned according to the measured shift.

Also, the $O$ and $E$ spectra have slightly different spectral
resolutions, again by a few tenths of a pixel, despite efforts during
focus set-up. We suspect that this is responsible for the erratic
behaviour in the polarization spectrum in unresolved spectral
features, such as a narrow P Cygni absorption components, which is
highlighted by comparing the different sets of an object. Degrading
the spectral resolution to make all spectra the same was $not$ found
to solve this problem. We suggest this is due to pixelation inherent
in the data due to the finite separation of pixels on the CCD. We
identified such artifacts by looking at the behaviour around narrow
features in each data-set, and have either edited them out through
linear interpolation or have been flagged in the text. This problem is
unique to studies of objects with narrow features such as the discrete
absorption components in LBV emission lines; it did not affect the
results of other objects studied in previous papers \citep [e.g.] []
{Oudmaijer99,Vink03}.

 \begin{figure}[!h]
 \begin{center}
\resizebox{\hsize}{!}{\includegraphics[bb=25 60 550 780,angle=0,clip]{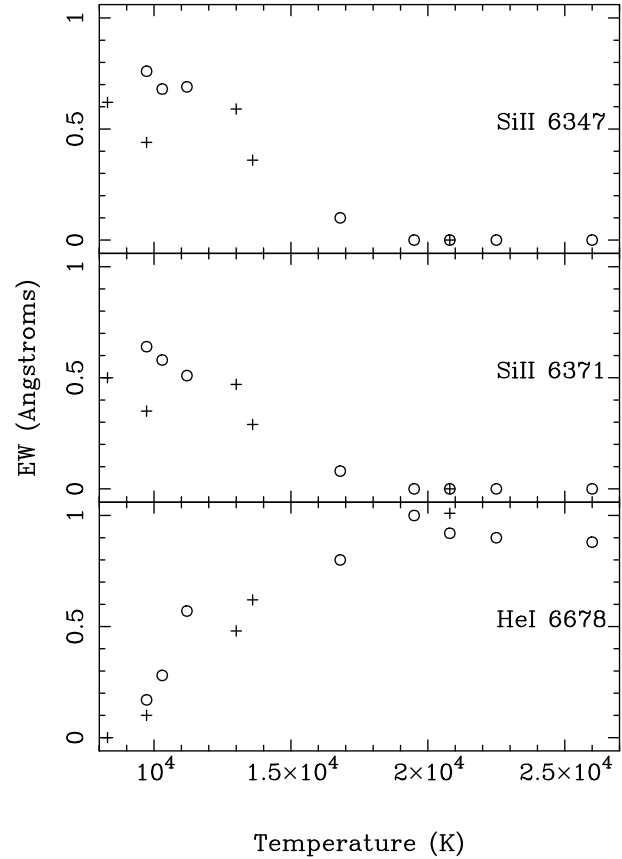}}
 \caption[]{The equivalent widths as a function of temperature for
 three photospheric absorption lines of early-type
 supergiants. Circles indicate stars with emission lines in their
 spectra, crosses stars with no emission. Note that the emission and
 non-emission stars follow the same relation. The uncertainties are of
 order the size of the plotting symbols and are not shown. The
 temperatures of the spectral types were obtained from
 \citet{S-K82}. \label{fig:ews}}
 \end{center}
\end{figure}

\section{Results \label{sec:results}}

The spectra of all the stars show strong H$\alpha$ emission, often
with P Cygni absorption components. Many show emission lines of Fe
\textsc{ii} and [N \textsc{ii}], whilst He \textsc{i} and Si
\textsc{ii} are seen in both emission and absorption. Of the objects
for which more than one set exists, only \object{AG Car} exhibits
significant spectral variability. The spectrum has changed from very
reminiscent of \object{HR Car} to one with emission lines only,
indicating a significant change in temperature and/or wind properties
between the two epochs.

In Section \ref{sec:obstemp} we estimate the temperature, and hence
phase (whether at minimum or maximum) at the epoch of each
observation. In Section \ref{sec:specpol} we display the results of
the spectropolarimetry and discuss each object individually.

\subsection{Temperature and Phase at Observation\label{sec:obstemp}}

The properties of line-driven stellar winds vary with effective
temperature for two reasons -- firstly, as a star cools a growing
mismatch is formed between its flux-peak and the bulk of the driving
lines in the UV \citep{L-C99}. Secondly, at around 21,000K the opacity
dramatically rises as the recombination of Fe \textsc{iv} to Fe
\textsc{iii} increases the number of opacity-enhancing ions \citep
[the so-called `bi-stability jump', see] [] {Vink99}. There is also
evidence for a second bi-stability jump at $\sim$ 10,000K, where Fe
\textsc{iii} recombines to Fe \textsc{ii} \citep{Lamers95,Vink99}.

As LBV effective temperatures can vary from 9,000K to up to 30,000K on
timescales of a few years, we may expect to find different wind
properties at different epochs. If the star's temperature is
latitude-dependent, due to e.g. rotation, we may expect to find
differential wind properties between equator and pole as the star
approaches its bi-stability jump, and therefore different wind
\emph{geometries} for the same star depending on the phase of the star
at observation. This mechanism has been used to explain the
two-component wind of B[e] supergiants by \citet{Pelupessy00}.  Here
we estimate the temperature of the star at observation using two
methods -- comparing the photospheric spectral absorption features
with an atlas of early-type supergiants, and via the stars' bolometric
correction from recent light-curves.

\begin {table}[]
 \begin{center}
  \begin {tabular}{lcc}
	\hline
 
Star& \multicolumn{2}{c}{Log $(T_{\rm eff})$}  \\
 &BC method&Abs. Line method \\
\hline \hline
\object{AG Car} (6/5/94)&3.98 $\pm$ 0.03&4.00 $\pm ^{0.04} _{0.05}$ \\
\vspace{1.5mm} \object{AG Car} (10/2/03)&4.25 $\pm$ 0.03&- \footnotemark[2] \\
\object{HR Car} (16/3/92)&3.87 $\pm$ 0.05& 3.98 $\pm ^{0.02} _{0.03}$ \\
\vspace{1.5mm} \object{HR Car} (10/2/03) & 3.92 $\pm$ 0.03 & 3.95 $\pm ^{0.04} _{0.05}$ \\
\object{$\eta$ Car}&-\footnotemark[1]  &- \footnotemark[2] \\
\object{Hen 3-519}& -\footnotemark[3]  &-\footnotemark[2]  \\
WRA 751&4.47 $\pm$ 0.03& 4.40 .. 4.52 \\
\object{P Cyg} & -\footnotemark[7]  &-\footnotemark[2]  \\
\object{HD 160529} &4.00 $\pm$ 0.03&3.95 $\pm ^{0.04} _{0.05}$ \\
R 40&4.04 $\pm$ 0.08&3.98 $\pm ^{0.03} _{0.04}$ \\
\object{S Dor} \footnotemark[4] & $\ga$ 3.93 & 3.95 $\pm ^{0.03} _{0.04}$ \\
R 71 & 4.04 $\pm$ 0.12 & 4.19 $\pm ^{0.03} _{0.04}$ \\
R 110 & 4.04 $\pm$ 0.08 \footnotemark[8] &4.00 $\pm ^{0.02} _{0.03}$ \\
R 116& 4.19 $\pm$ 0.04 & $\ga$ 4.15 \footnotemark[6]\\
R 127&4.29 $\pm$ 0.04&4.13 $\pm ^{0.08} _{0.11}$ \\
R 143& -\footnotemark[5]  &3.99 $\pm ^{0.04} _{0.05}$ \\

      \hline
      \end {tabular}
     \end{center}

\caption{Effective temperature at observation as calculated from the BC and spectral characteristics methods (see text for details). The two estimates agree well for most of the objects, exceptions are explained below.  \label{tab:obstemps}}

\hspace{2mm}$^{1}$\hspace{1mm}The poor understanding of the intrinsic brightness of \object{$\eta$ Car} means that the BC method does not produce a reliable result. \par
\hspace{2mm}$^{2}$\hspace{1mm}The star's spectrum does not contain the diagnostic absorption lines. \par
\hspace{2mm}$^{3}$\hspace{1mm}An insufficient amount of data is available on the star's brightness history and brightness at the time of observation. \par
\hspace{2mm}$^{4}$\hspace{1mm}There is disagreement on the star's temperature at minimum and maximum (as stated in vG01), giving two possible values of the star's temperature at observation using the BC method. Also the star's bolometric luminosity probably changes during outbursts (vG01), invalidating the BC method. \par
\hspace{2mm}$^{5}$\hspace{1mm}No AAVSO data exists on this object. \par
\hspace{2mm}$^{6}$\hspace{1mm}The Si \textsc{ii} absorption lines appear to be filled in by emission in this object, so the temperature was determined from the He \textsc{i} line only. \par
\hspace{2mm}$^{7}$\hspace{1mm}\object{P Cyg} is not presently observed to be variable, therefore the BC method is not applicable. \par
\hspace{2mm}$^{8}$\hspace{1mm}R 110 probably does not vary at constant $L_{\rm BOL}$, and so the BC method is invalid (vG01). \par
\vspace{3mm}
\end {table}

\subsubsection{A Spectral Atlas of Early-Type Supergiants}

The spectra of emission line and non-emission line early-type
supergiants in the range 6240 -- 6700\AA \space can be found in
Appendix \ref{sec:app}, and the full data is available electronically
at {\tt http://vizier.u-strasbg.fr/viz-bin/VizieR}. Among the various
spectral features, three lines in particular display a strong
temperature dependence -- the Si \textsc{ii} $\lambda \lambda$6347,
6371 lines and the He \textsc{i} $\lambda$6678 line, whilst no
difference is observed between the emission and non-emission line
stars. Figure \ref{fig:ews} shows the measured equivalent widths (EWs)
of these lines as a function of effective temperature. The EW of the
He \textsc{i} line clearly increases with increasing temperature,
appearing at around 8000K (spectral-type A7) and peaks at 19 --
21,000K (B1 -- B1.5), in agreement with \citet{S-K82} who states that
He \textsc{i} peaks at spectral type B0. The Si \textsc{ii} lines
decrease in EW with temperature, peaking at 9 -- 10,000K (B9 - A2) and
disappearing at around 19,000K (B2). This agrees well with
\citet{S-K82}, who states that Si \textsc{ii} lines peak at
spectral-type AO (9700 K). We find no difference between emission and
non-emission stars. It appears, therefore, that these three lines are
reasonable indicators of effective stellar temperature in the range
9,000K -- 20,000K ($\sim$ A2 -- B1). By comparing with the spectral
features of the LBVs we can obtain an empirical estimate of $T_{\rm
eff}$, and therefore of the phase of the LBV at observation. This
empirical relation will break down if the lines are in emission, but
will only give a misleading temperature if there is a tiny amount of
emission which only partially fills in the absorption.

\subsubsection{The Bolometric Correction Method}

The light variations of LBVs occur at approximately constant
bolometric luminosity \citep[][ VG01 hereafter]{vG01}. Their apparent
variability is caused by their spectral energy distribution (SED)
shifting in and out of the visual band as their effective temperature
changes. If we know the apparent bolometric magnitude, we can use the
difference between this and the apparent magnitude at observation (the
bolometric correction, BC) to determine the star's temperature. This
method has previously been used to determine LBV temperatures by
\citet{Lamers98}.

The BC as a function of $T_{\rm eff}$ for supergiants is quoted in
\cite{S-K82}. We have fitted this to within 0.05mags with the
following relations:

\begin{subequations}
\begin{equation}
BC = +27.22 - 6.74 \log T_{\rm eff}, \space (\log T_{\rm eff} > 4.25) \label{equ:bc1}
\end{equation}

\begin{equation}
BC = +15.47 - 3.98 \log T_{\rm eff}, \space (\log T_{\rm eff} < 4.25) \label{equ:bc2}
\end{equation}
\end{subequations}

VG01 quotes the $m_{\rm V}$ at visual minimum, with the associated
$T_{\rm eff}$, for each LBV as collated from multiple sources. From
this, the BC and hence $m_{\rm Bol}$ can be calculated. The $m_{V}$ at
observation was obtained from the American Association of Variable
Star Observers (AAVSO, http://www.aavso.org/; Waagen 2005), and the BC
at the epoch of observation was calculated from the difference between
this and $m_{\rm Bol}$. Equations (\ref{equ:bc1}) and (\ref{equ:bc2})
were then used to calculate the $T_{\rm eff}$ at observation.

\begin{figure*}[tb]
\sidecaption
 \includegraphics[width=10cm,angle=90]{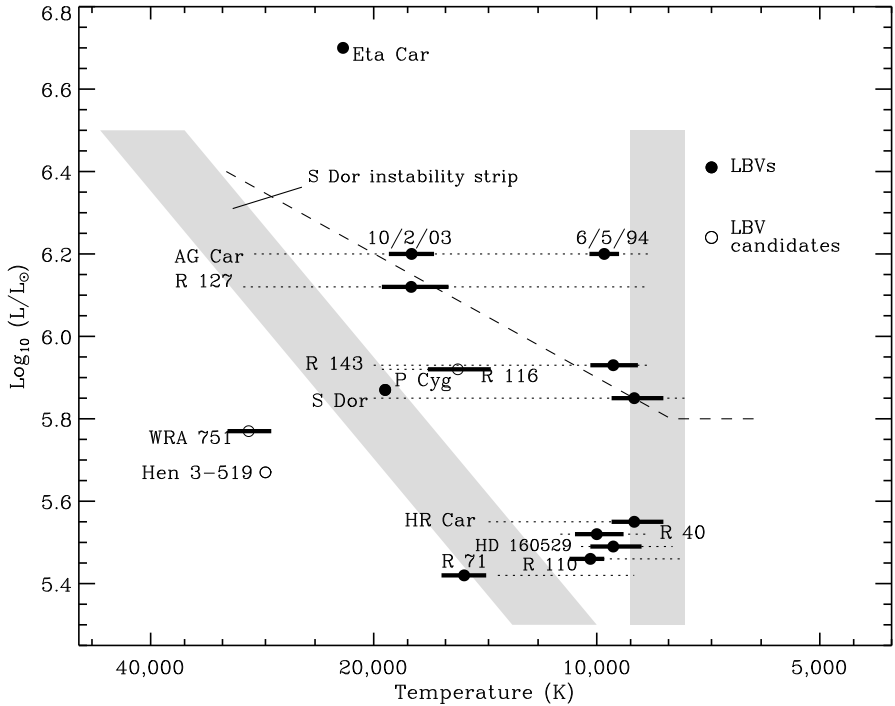}
 \caption[]{HR diagram illustrating the temperature of each LBV at the
 epoch of observation. Filled and open circles show the temperature
 determined at observation for LBVs and candidate LBVs respectively,
 with the solid lines showing the error in each measurement. The
 luminosities, and minimum and maximum observed temperatures were
 collated from the literature by \cite{Smith04}. The
 Humphreys-Davidson limit is marked by a dashed line. \object{$\eta$
 Car}, Hen 3-519 and \object{P Cyg} are marked on the plot despite the
 fact that their temperatures could not be determined -- values are
 taken from Smith et al. and \citet{Cox95}.
 \label{fig:hrdiag}}
\end{figure*}

The estimations of effective temperature at observation are shown in
Table \ref{tab:obstemps}. The two independent measures of temperature
agree well for each object. Exceptions are marked in the table, and
probable explanations given in the footnotes. Figure \ref{fig:hrdiag}
illustrates the phase of each LBV at observation by placing them on a
HR diagram, along with their locations at visual minimum and maximum.

\subsection{Wind Geometry \label{sec:specpol}}

As explained in Section \ref{sec:intro}, changes in polarization
across emission lines supply evidence for aspherical stellar
winds. The strength of the change in polarization is dependent on both
the density contrast between different regions in the wind and the
inclination angle. A non-detection could imply either the wind is
symmetric about some axis (e.g. an equatorially-enhanced wind) but is
oriented face-on; or the electron density at the base of the wind is
isotropic (i.e. the wind is spherically symmetric) to within the
detection limit. This detection limit is dependent inversely on the
S/N of the spectrum, and the contrast of the emission line as the
line-emission is `diluting' the polarized flux from the continuum. The
detection limit per pixel for the maximum intrinsic polarization
$\Delta P_{int,max}$ for each object is given by:

\begin{equation}
\Delta P_{\rm int,max} (\%) = \frac{1}{S/N} \times \frac{L}{L-1} \times 100
\label{eq:pint}
\end{equation} 

\noindent where $L$ is the line-to-continuum contrast. In the cases
where no line-effect is observed, we state the systematic detection
limit of our observations. Where line-effects $are$ detected, the PA
of the intrinsic component of the star's observed polarization is
determined from the line-to-continuum vector in $Q-U$ space using
Eqns. (\ref{equ:p}) and (\ref{equ:th}). Note that this is independent
of the interstellar polarization (see below).

The continuum polarization is discussed below, and properties of the
H$\alpha$ emission as well as continuum polarization measurements are
shown in Table \ref{tab:Hadata}. Object-by-object descriptions are
listed below, starting with the Galactic objects. Of the 14 targets
observed, 6 show definite line-effects, with one further possible
detection. Of the 4 objects for which more than one observation exist,
3 show line-effects, and each of these three display polarimetric
variability.

\subsubsection{Continuum Polarization \label{sec:contpol}}

The observed polarization in the continuum is a superposition of
multiple sources. We assume that this is broadly due to three
mechanisms: scattering of light by free electrons at the base of the
stellar wind; scattering by gas and dust in the star's nebula; and
dichroic absorption by dust in the interstellar medium (ISM). Here we
focus on the polarizing effects of asphericities in the stellar wind
within a couple of stellar radii, which can vary on short timescales
(days). For the purposes of this paper we will lump the nebula and the
ISM together, and say the continuum polarization is due to the
interstellar polarization (ISP) and electron scattering in the stellar
wind.

Determination of the ISP, and hence separation of the two polarizing
components is not straight-forward. The field-star method, attempted
by \citet{S-L94,Harries98,Parth00} amongst others, generally yields
mixed results due to uncertain distances, the angular scale on which
the ISP can change, and the unknown intrinsic polarization of the
field stars. The line-centre method, used by
\citet{S-L93,S-L94,Clampin95,Oudmaijer98}, assumes that as the
H$\alpha$ line-emission is intrinsically unpolarized and formed over a
much larger volume it will undergo negligible scattering and remain
unpolarized. Therefore, if the line is much stronger than the
continuum (such that the line contrast term in Eqn. (\ref{eq:pint})
tends to 1), the H$\alpha$ line-centre polarization is a measure of
the ISP. Multiple Spectropolarimetric measurements of \object{AG Car} by
\citet{S-L94} showed that the polarization of H$\alpha$ remained
constant over $\sim$ 1.5 years whilst the continuum polarization
varied, showing that this method can be an effective method of
determining the ISP.

\begin{table*}[htb]
 \begin{center}
  \begin {tabular}{lcccccc}
	\hline
      	Name 	& Line-   & Line-cont. & EW    & FWHM  &
      	\multicolumn{2}{c}{Cont. Pol. ($R$-band)} \\ 
			& Effect? & Contrast       & (\AA) & (\kms)  & P (\%) & $\theta$ (\deg) \\
      	\hline \hline
	\emph{Galactic} & & & & & \\
	\object{AG Car} \tiny{(6/5/94)}& Y	  & 17.5	& -80 & 117& 0.95 $\pm$ 0.01 & 160.1 $\pm$ 0.2  \\	  
\vspace{2mm}'' \tiny{(10/2/03)}	& Y 	  & 25	& -119  & 150  & 1.09 $\pm$ 0.01 & 132.6 $\pm$ 0.4  \\

	\object{HR Car} \tiny{(16/3/92)}	& Y 	  & 16	& -59   & 115 & 4.22 $\pm$ 0.01 & 129.7 $\pm$ 0.1 \\
\vspace{2mm} '' \tiny{(10/2/03)}	& Y 	  & 18	& -90   & 95  & 3.37 $\pm$ 0.01 & 133.9 $\pm$ 0.1   \\

	\object{$\eta$ Car}	& Y	  & 100	&	-1480 & 450 & 5.71 $\pm$ 0.04 & 76.4 $\pm$ 0.2  \\

      \object{Hen 3-519} \tiny{(6/5/94)}	& N	  & 17	& -197 & 475 & 2.90 $\pm$ 0.02 & 97.6 $\pm$ 0.2 \\
\vspace{2mm} '' \tiny{(10/2/03)}	& Y  	  & 22		  	      & -220  & 435  & 3.05 $\pm$ 0.01 & 99.2 $\pm$ 0.1  \\

	WRA 751	& N  	  & 4.2		         & -10   & 90 & 2.29 $\pm$ 0.01 & 58.7 $\pm$ 0.1  \\
	\object{P Cyg} \tiny{(10/12/03)}	& N 	  & 18		      & -63   & 167 & 0.99 $\pm$ 0.02 & 29.3 $\pm$ 0.4 \\
\vspace{2mm}
	''    \tiny{(13/12/03)}	& N 	  & 18		  	      & -64   & 170  & 0.93 $\pm$ 0.02 & 29.3 $\pm$ 0.6 \\
\vspace{2mm}	\object{HD 160529} 	& N 	  & 2		   	      & -4.4  & 80& 6.97 $\pm$ 0.01 & 21.2 $\pm$ 0.1 	\\

	\emph{SMC} & &&&&& \\

\vspace{2mm}	R 40	& N	  & 4.6		      & -15   & 87& 0.28 $\pm$ 0.03 & 92.9 $\pm$ 2.6 	\\
	\emph{LMC} &		&	&	&	&& \\
	\object{S Dor}	& ?	  & 8.3		  	      & -34   & 100 & 0.24 $\pm$ 0.02 & 72.6 $\pm$ 2.6 \\
	R 71  & N	  & 2.0 	        & -7.8  & 200 & 0.31 $\pm$ 0.03 & 60.1 $\pm$ 2.3 \\	
	R 110 & Y	  & 26		  	      & -96   & 100 & 1.50 $\pm$ 0.03 & 32.1 $\pm$ 0.6 \\
	R 116 & N	  & 2.5			      & -6.6  & 130 & 0.51 $\pm$ 0.02 & 28.6 $\pm$ 1.0 \\
	R 127 & Y	  & 40			      & -114  & 78 & 0.50 $\pm$ 0.01 & 33.4 $\pm$ 0.5 \\
	R 143 & N	  & 6.4		 	      & -20   & 100 & 2.93 $\pm$ 0.02 & 73.3 $\pm$ 0.2 \\
	
      \hline
      \end {tabular}
     \end{center}
\caption{The observed H$\alpha$ data. The first column denotes whether
or not a line-effect was detected; the second column shows the
contrast between the line-peak and the continuum to an accuracy of
0.1; the equivalent-widths and FWHM are measured to an accuracy of
$\sim$ 5\%. The continuum polarization was measured from featureless
sections either side of the emission line. As the R 40 data from
consecutive nights showed no variability, the data was combined to
improve the S/N. The quoted uncertainties do not include the external
errors, which we estimate to be 0.1 -- 0.2\% (see Section
\ref{sec:obs}).\label{tab:Hadata}}

\end {table*}

\subsubsection{Object-by-Object Descriptions}

\paragraph{AG Car} 
One of the prototype LBVs, \object{AG Car} has been highly variable
for over 30 years with light variations of over 2 mags.  It is
surrounded by an elliptical nebula which has a hint of a bi-polar
morphology, and a dynamical age of $\sim 10^{4}$ years \citep{Nota92}.
High-resolution HST images have shown that the nebula is made up of
individual clumps, filaments and bubbles \citep{Nota95}.  Multiple
spectropolarimetric observations revealed evidence for an axisymmetric
stellar wind, with a PA alternating between perpendicular and parallel
to the major-axis of the nebula \citep[][ SL94 hereafter]{S-L94}. SL94
attributed this to either a wind which flips between an equatorial-
and polar-enhanced wind, or multiple-scattering effects due to changes
in optical depth.

\begin{figure*}[htb]
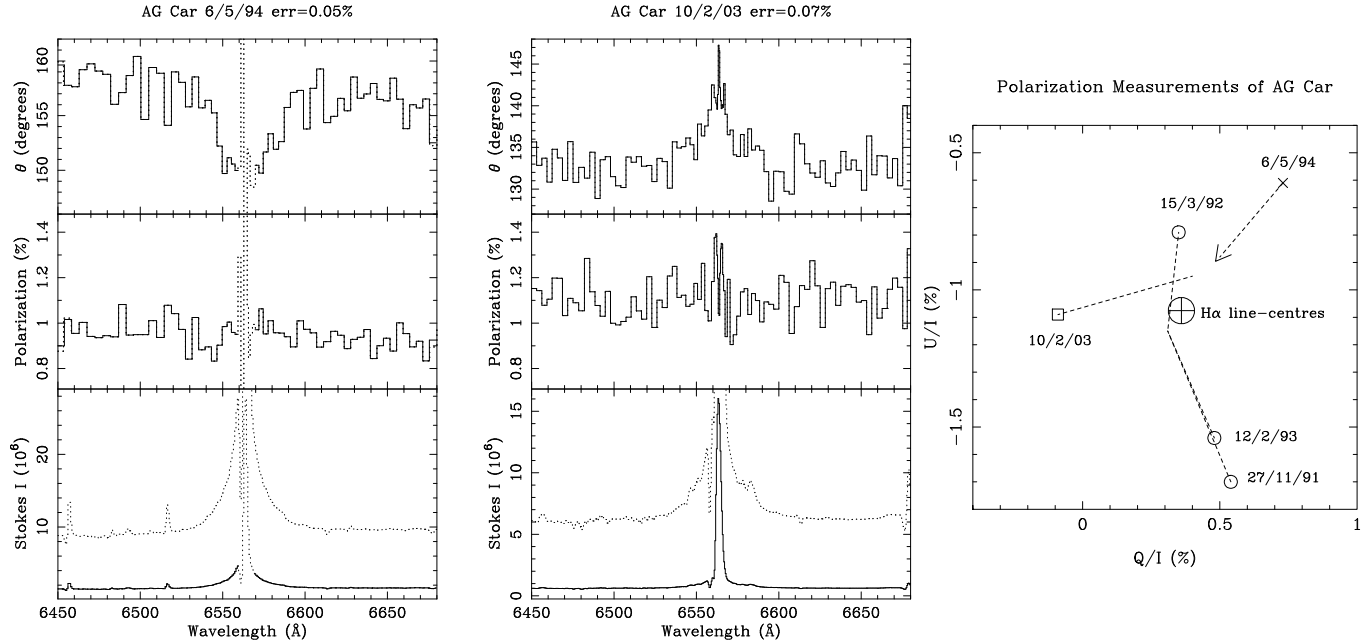

\includegraphics[width=12cm,clip]{2781fg3a.ps}
\includegraphics[bb=-10 30 500 705,width=6cm,clip]{2781fg3b.eps}
 \caption[]{\emph{Left, Centre}: Polarization spectra of \object{AG
 Car} from the two dates observed. The bottom panel of each triplot
 shows the Stokes $I$, or intensity spectrum; and the middle and top
 panels show the degree and PA of the polarization respectively as a
 function of wavelength. The data has been rebinned to the 1$\sigma$
 error indicated above each plot. The 6/5/94 data displayed erratic
 behaviour around the absorption component and line-centre due to
 reasons explained in Sect. \ref{sec:obs}, and was edited out of this
 plot for clarity.

\emph{Right}: Multiple spectropolarimetric measurements of \object{AG
Car} plotted in $Q-U$ space. Circles are continuum points from SL94;
the cross is from archival data; and the square is our continuum
data. Dotted lines show the excursions to the H$\alpha$ line-centre
polarization. Note that the 1994 excursion was attenuated due to
fluctuating data (see Section on \object{AG Car}), and that SL94
averaged their three line-centre polarizations together. The mean of
the line-centre polarizations is marked on the plot, with a 1$\sigma$
rms error of the size of the symbol. Uncertainties in the continuum
are limited by the external errors, of order 0.1\%. }
\label{fig:agcar}
\end{figure*}

The data from both epochs show line-effects (see Fig.
\ref{fig:agcar}). The archive data from 1994 had problems with varying
spectral resolutions from set to set. Consequently, erratic
polarization behaviour around the emission line was observed, and has
been edited out in order to show the broad PA rotation coincident with
the broad wings. Similar behaviour is seen in the 2003 data, but the
continuum PA has changed by $\sim$ 30\deg. The varying PA of the
observed polarization alone is evidence for an intrinsic component, as
the interstellar polarization (ISP) will not vary on these timescales
(see Sect. \ref{sec:contpol}).

The line-centre polarization of \object{AG Car} has previously been
shown to remain roughly constant, and is therefore probably a good
estimate of the ISP (SL94, see Sect. \ref{sec:contpol}). As shown in
Fig.  \ref{fig:agcar}, the ISP measured by SL94 agrees well with the
line-centre polarization of our 2003 data. The full $Q-U$ excursion of
the 1994 data has been attenuated due to the line-centre being edited
out, but it can clearly be seen to point towards the same region of
$Q-U$ space. The mean of the line-centre polarizations, including
those measured by SL94, is $Q = 0.36 \pm 0.07; U = -1.08 \pm 0.09$.

Due to the attenuation of the 1994 data, the intrinsic polarization is
measured from the line-wings, and is found to be $P > 0.26 \pm 0.03\%,
\theta = 26 \pm 3 \deg$. The intrinsic polarization of the 2003 data
is $P = 0.41 \pm 0.03\%, \theta = 84 \pm 2\deg$. A weak line-effect is
also seen across the He \textsc{i} $\lambda$6678 line which is in
emission at this epoch, with a similar PA to H$\alpha$. However,
line-effects are not observed across any other emission lines. The PAs
from both epochs, 26\deg ~and 84\deg, appear to be aligned with
neither the nebula ($\sim$ 135\deg), nor the jet ($\sim$ 35\deg).

Looking at our data and the measurements of SL94, it can be seen that
the intrinsic polarization of \object{AG Car} varies greatly from
epoch to epoch (Fig. \ref{fig:agcar}). The data from SL94 showed the
intrinsic polarization vector flipping from one side of the origin in
$Q-U$ space to the other, leading them to suggest that the intrinsic
polarization varied between two preferred planes 90\deg \space to each
other. These planes aligned with the major and minor axes of the
coronographic image of the \object{AG Car} nebula in
\citet{Clampin93}. Our observations lie in neither of these planes,
with the archive data being taken only $\sim$ 6 months after the last
SL94 observation.  The continuum measurements of \citet[ not shown in
Fig. \ref{fig:agcar}]{Leitherer94} are different again. If the wind
was indeed flipping from an equatorial to a polar flow, we would
expect the polarization of any intermediate phases to lie along the
same axis in $Q-U$ space but with a smaller line-to-continuum
vector. This is not observed here, with strong depolarizations
observed at all epochs at unrelated PAs. Unless the assumption used to
estimate the ISP are invalid, the evidence points away from the
flip-flopping axi-symmetric wind scenario. Instead, it may point
towards a `clumpy' wind model, where the PA of the intrinsic
polarization at any time is related to the projected angle between the
star and the dominant clump(s). This is the same scenario to that
proposed for \object{P Cyg} by \citet[][ see section on \object{P
Cyg}]{Nordsieck01}.

\paragraph{HR Car} 
Like \object{AG Car}, \object{HR Car} has been strongly variable for
over 25 years ($\ga$ 1 mag on timescales of $\sim$ 3 years). It is
surrounded by a filamentary bi-polar nebula (PA $\sim$ 135\deg) which
has a kinematic age of $\sim$ 5000 years \citep{Nota97}. The star was
noted by \cite{Serk75} as polarimetrically variable, implying it has
intrinsic polarization. This was confirmed by subsequent polarimetric
monitoring by \citet{Parth00}. It was observed spectropolarimetrically
by \citet{Clampin95}, who found evidence for intrinsic polarization at
a PA of 30 $\pm$ 1\deg, which is aligned with the `finger' of nebular
emission (PA $\sim$ 35\deg) within $\sim$ 3\arcsec \space of the
central star noted in their coronagraphic image. They also noted the
presence of a number of clumps of emission close to the star, which
did not correspond to the filamentary structures of the nebula and
appeared to be roughly spherically distributed around the star.

\begin{figure*}[htb]
\includegraphics[bb=20 30 390 300,width=12cm,clip]{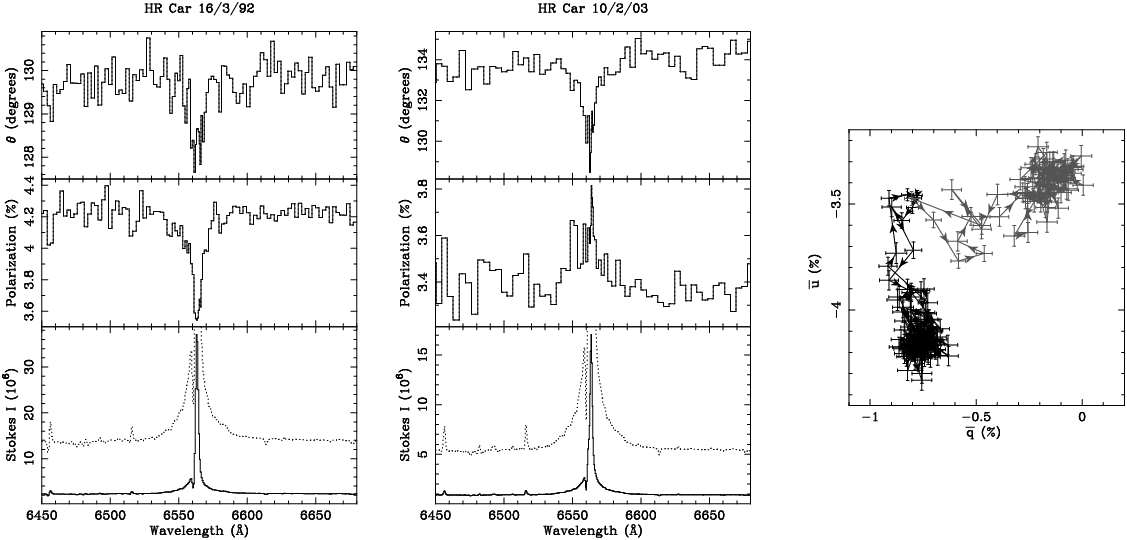}
\includegraphics[bb=-10 50 500 700,width=6cm,clip]{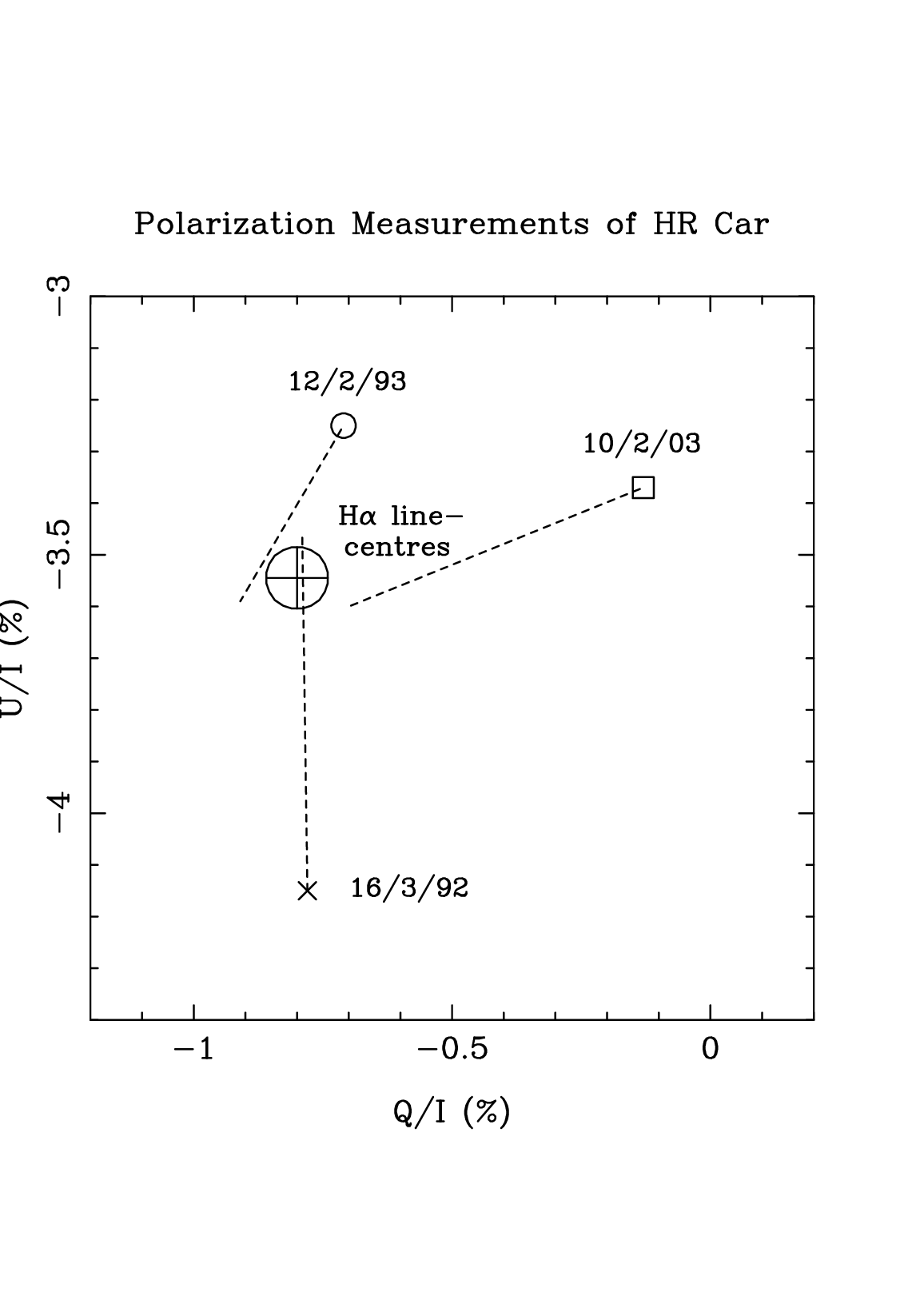} 
 \caption[]{\emph{Left, Centre}: Polarization spectra of \object{HR Car} from the two dates observed. The panels show the same as Figure \ref{fig:agcar}, and the data has been rebinned to a 1$\sigma$ error of 0.05\%. 

\emph{Right}: Multiple spectropolarimetric measurements of \object{HR Car},
plotted in $Q-U$ space. The cross marks the continuum polarization at
H$\alpha$ for the 1992 archival data; the circle for the data taken by
\cite{Clampin95}; and the square for our data. The dotted lines show
the excursions to the line-centre polarization. The mean of the
line-centre polarizations is marked, with a 1$\sigma$ rms error of the
size of the symbol. Uncertainties in the continuum are limited by the
external errors, of order 0.1 -- 0.2\%.}
\label{fig:hrcar}
\end{figure*}

In our spectropolarimetry, both epochs show line effects (see Fig.
\ref{fig:hrcar}). The 1992 data shows a depolarization accompanied by
a PA rotation. The 1992 intrinsic polarization is $P = 0.67 \pm 0.03
\%, ~\theta = 138 \pm 2 \deg$, which is aligned parallel to the major
axis of the bi-polar nebula. The 2003 data shows a lower continuum
polarization, with an increase in polarization across the emission
line. The 2003 intrinsic polarization is $P = 0.56 \pm 0.03 \%,
~\theta = 13 \pm 2 \deg$. This PA bears no obvious relation to the PA
measured at the previous epoch, nor to any large-scale nebular
features.

The line-centre polarization from our observation, the archival data
and the data in \citet{Clampin95} agree well with each other,
indicating that this is a good estimate of the ISP towards \object{HR
Car} at 6563\AA\ (as with \object{AG Car}). The mean of these
measurements is $Q = -0.80 \pm 0.05 \%, ~U = -3.56 \pm 0.03 \% $;
implying $P = 3.66 \pm 0.05 \%, ~\theta = 128.7 \pm 0.4 \deg$ (see
Figure \ref{fig:hrcar}, \emph{right}).

Line-effects are also seen across Fe \textsc{ii} lines at both
epochs. In the 1992 data, there is a borderline detection across the
$\lambda$6456 line, with a line-to-continuum PA of $\sim$ 140\deg,
i.e. parallel to the H$\alpha$ depolarization. In the 2003 data,
line-effects are detected across the Fe \textsc{ii} ($\lambda \lambda$
6417, 6433, 6456, 6516) lines. The depolarizations all have PAs
between 35 -- 60\deg, but these are borderline detections.

Taking the ISP to be that calculated above we find that the Clampin
data and archival data, taken approximately 1 year apart, are
perpendicular to each other with PAs of $\sim 45 \deg$ and $\sim
135\deg$ respectively. These align with the major and minor axes of
the bipolar nebula. Our data, taken 10 years later, has a completely
different PA of $\sim 15 \deg$. Taking the spectropolarimetry of
\citet{Clampin95} and multiple $R$-band polarimetric measurements of
\citet{Parth00} into account, the intrinsic polarization of \object{HR
Car} seems to have no preferred axis with time. In a clumpy wind we
may expect this kind of temporal variability, where the polarization
at each epoch is related to the projected PA between the star and the
closest, densest clumps.

\paragraph{$\eta$ Car} 
The most famous of the LBVs underwent a huge eruption in the mid-19th
century when it increased in brightness by over 2 mags and ejected
$\sim$ 10$\msun$ \citep{Morris99,Smith03aj}, the product of which is
the `homunculus' nebula we see today \citep[for an extensive review of
this object see][]{D-H97}. Aside from a smaller eruption in the late
19th century, which is possibly responsible for the mini-homunculus
within the main nebula detected by \citet{Ishi03}, the star has been
more or less stable. The periodic ($\sim$ 2000 days) spectroscopic
events are attributed to the star being in a highly eccentric binary
system \citep{Damineli00}; and the gradual brightening in the optical
and NIR is attributed to the expansion of the nebula
\citep{D-H97}. \object{$\eta$ Car}'s wind as a function of latitude
was studied by \citet{Smith03apj} by looking at starlight reflected of
the surrounding nebula. They determined that the wind has higher
terminal velocities and higher mass-fluxes towards the poles, assuming
the star's axis is aligned with the bipolar nebula. Inteferometric
observations by \citet{vB03} revealed that the star's NIR emission is
prolate, with a major axis of PA $134 \pm 7 \deg$. This is aligned
with the PA of the humunculous nebula \citep[PA 132\deg,][]{Morse98}.

\begin{figure}[htb]
\centering
\includegraphics[width=7cm]{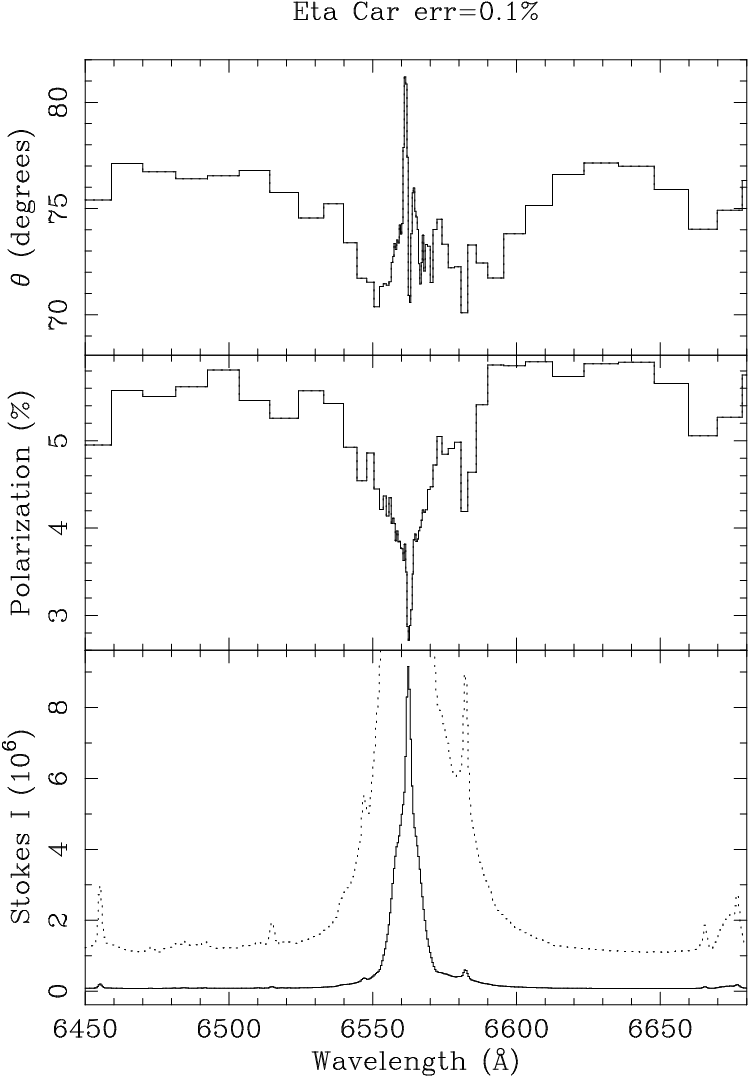}
\includegraphics[width=7cm]{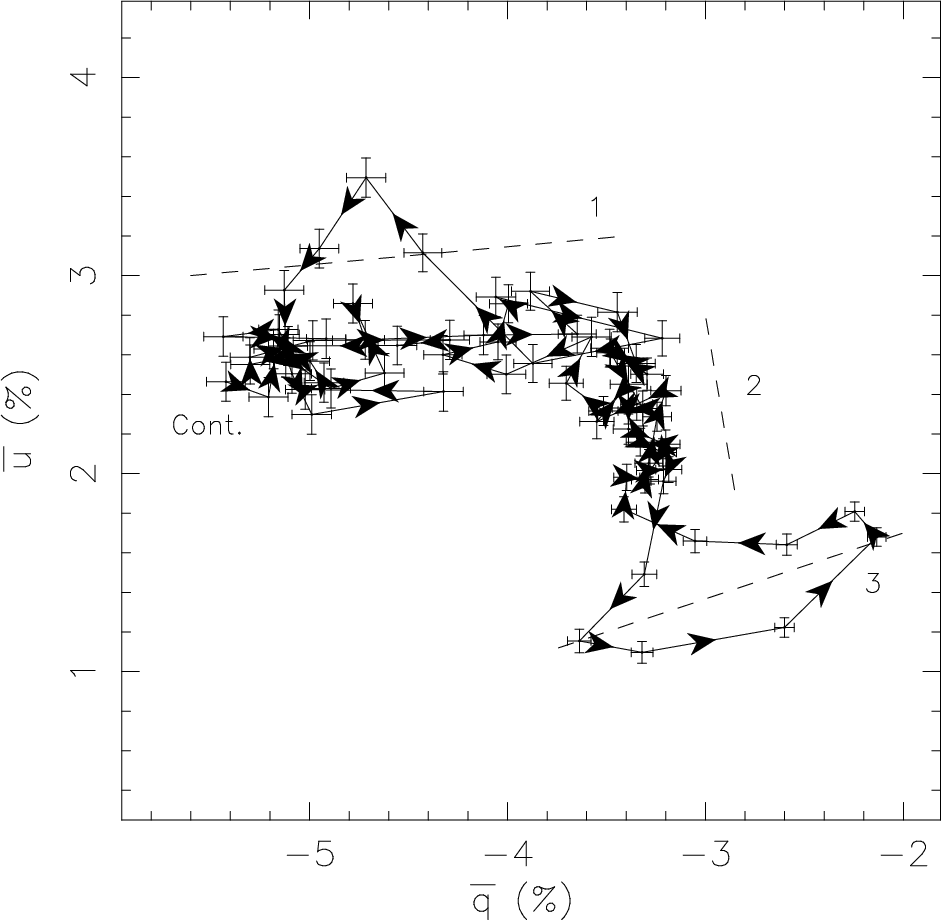}
 \caption[]{\emph{Top}: The polarization spectrum of \object{$\eta$ Car}. The
 panels show the same as Figure \ref{fig:agcar}. The data has been
 rebinned to a 1$\sigma$ error of 0.1\%. The dotted line shows a
 magnification of the Stokes $I$ spectrum to highlight the broad wings
 and weaker spectral lines.
 
\emph{Bottom}: The spectrum plotted in $Q-U$ space. The arrows join
points of increasing wavelength. The location of the continuum is
marked `$cont.$'. The distinct excursions mentioned in the text are
indicated with dotted lines. The excursion associated with the wings
is marked `$1$'; the excursion associated with the narrower region
containing the bulk of the line-flux is marked `$2$'; and the loop
associated with the narrow, unresolved nebular emission is marked
`$3$'.}
\label{fig:etacar}
\end{figure}

Our data shows a very complex polarization profile, corresponding to
an emission line which is made up of three components -- the broad
wings, the narrower component containing the bulk of the flux, and the
narrow peak (see Fig. \ref{fig:etacar}). The broad wings correspond
to the $Q-U$ excursion labelled $1$ with a PA of 91\deg; the flux-bulk
to the excursion labelled $2$ with a PA of 52\deg; and the narrow peak
to the $Q-U$ loop marked $3$, with a PA of 95\deg. There is also a
loop as the vectors return from the high-flux region to the continuum
region. This does not correspond to any detectable spectral features.

Comparison of our spectrum with archive data taken by the Hubble Space
Telescope (\emph{HST}) one day previously reveals that the narrow
emission peak (\#3) does not originate from the star, but from a
nebulous region approximately 1\arcsec ~away. Our data does not
spatially resolve the star and the nebula, and so this nebulous
emission will not have been removed at the sky-subtraction stage of
data reduction (see Sect. \ref{sec:obs}). The [N \nolinebreak
\textsc{ii}] emission line redward of the H$\alpha$ line, which
displays a depolarization, also originates from the nebula. As the
nebula is located well outside the polarizing region close to the
star, any polarization it may have will be unrelated to that of the
continuum. Closer inspection of the {\em HST} spectrum reveals that
the nebula also has continuum emission, probably from reflected
starlight.

Given that the light we see in our spectrum is made up of a direct
starlight component and a reflected nebula component, these may be
responsible for the distinct $Q-U$ excursions. The emission component
containing the bulk of the flux (`$2$' on Fig. \ref{fig:etacar},
which probably corresponds to a line-forming region close to the star)
with a PA of $52 \pm 3\deg$ is perpendicular to the major-axis of the
resolved prolate structure detected by \citet{vB03}. This is
consistent with polarization due to scattering off a bipolar wind, as
per the scenario suggested by \citet{Smith03apj}. The level of
depolarization, at $\sim$ 1\%, is also consistent with such a scenario
\citep{Cassinelli87}. The other $Q-U$ excursions, which seem to have
PAs of $\sim$ 95\deg \space and a much stronger depolarization of
$\ga$ 2\%, may be due to scattered light off the unresolved nebulous
component. This hypothesis is supported by other sharp features of the
spectrum (not shown here), which also show depolarizations with
similar PAs of $\sim$ 90\deg. These PAs do not relate to any nebular
features nor to the PA of the slit, but as shown in \citet{S-L99} the
polarization of \object{$\eta$ Car} varies greatly within 2\arcsec ~of the
central source.

In summary, the complex polarization profile may be a combination of
spatially unresolved emission from the nearby reflection nebula and
emission from the stellar wind. The polarization of the light from the
stellar wind is perpendicular to the bipolar axis of the homunculus
nebula. This is consistent with single scattering off an optically
thin bipolar wind, or multiple scattering off an optically thin
equatorially-enhanced wind \citep{Angel69,Wood96}.

\paragraph{Hen 3-519} 
This object is classified as a post-LBV/pre-WR star by
\citep{Davidson93}. It is not observed to be variable and has broad
emission-lines in its spectrum. For these reasons it is not possible
to determine the star's temperature at observation using the two
methods outlined in Sect. \ref{sec:obstemp}.  The star is surrounded
by a roughly circular nebula of diameter $\sim$ 55\arcsec. The nebula
appears to be a clumpy bubble, rather than a detached ring as seen
around \object{AG Car} \citep{Stahl87,D-W02}. The spectra from the two
different epochs separated by $\sim$ 9 years look very similar, with
the exception of a P Cygni profile present in the H$\alpha$ line of
the Feb 2002 data not present in the 1994 data.

\begin{figure*}[htb]
\includegraphics[width=17cm]{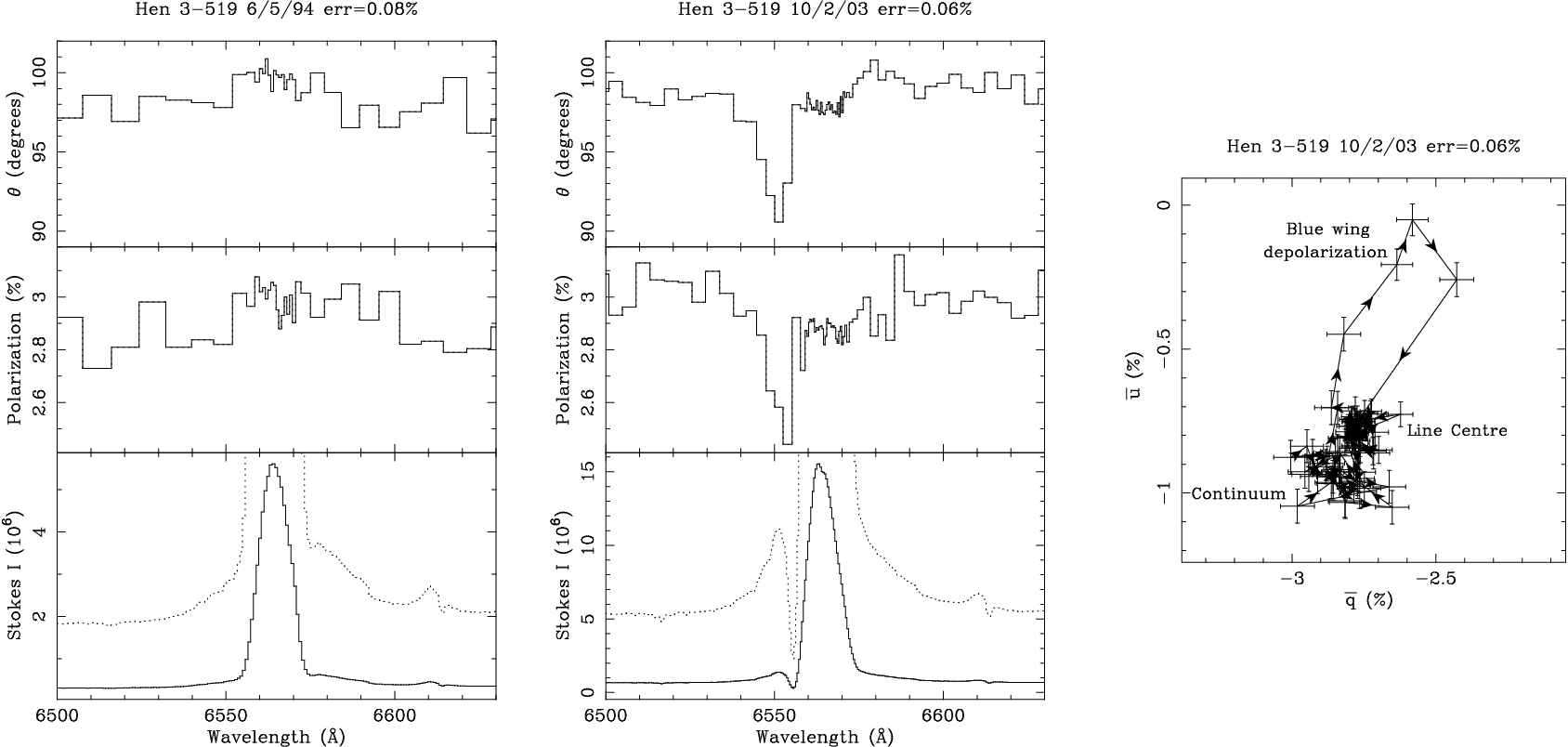}
 \caption[]{\emph{Left, Centre}: The polarization spectra of \object{Hen 3-519}
 from 6/5/94 and 10/2/03 repectively. The panels show the same as
 Figure \ref{fig:agcar} and the data has been rebinned to the
 1$\sigma$ error indicated above each plot.

\emph{Right}: The polarization spectrum of the 10/2/03 data plotted in
$Q-U$ space. Arrows join points of increasing wavelength. The
distinctly separate continuum and line-centre regions have been
marked, as well as the loop corresponding to the depolarization seen
in the blue wing.  }
\label{fig:he3519}
\end{figure*}

The archive data from 6/5/94 (Figure \ref{fig:he3519}, \emph{left})
does not show an obvious line-effect. The data from 10/2/03 shows
different polarization behaviour across the features of the line (see
Fig. \ref{fig:he3519}, \emph{centre}). The line-centre polarization
is consistent with that of the archival data, showing up in $Q-U$
space as a dark cluster of points above the continuum region. This
behaviour is also seen in the HeI $\lambda \lambda$ 5876, 6678
lines. The mean line-to-continuum vector, and hence intrinsic
polarization, is $Q = -0.10 \pm 0.03 \%$, $U = -0.14 \pm 0.03 \%$; $P
= 0.17 \pm 0.04 \%$, $\theta = 117 \pm 12 \deg$. The mean of the
line-centre polarizations is $P = 2.89 \pm 0.02 \%$, $\theta = 97.9
\pm 0.3 \deg$, which agrees with the data of the previous epoch.

As well as the H$\alpha$ line having a slightly different polarization
to the continuum, there is a large change in polarization in the blue
wing of the 2003 data, showing up as a looped excursion in $Q-U$ space
(see Fig. \ref{fig:he3519}, $right$). This behaviour in the
blue-wing is also seen in the He \textsc{i} $\lambda \lambda$ 5876,
6678 lines.  The mean change in polarization from the line-centre
polarization of these loops is $P = 0.87 \pm 0.08 \%$, $\theta = 25
\pm 3 \deg$.  At present, we have no explanation for these features,
which cannot be due to the simple dilution of the polarized continuum
photons. They are also unlikely to be related to the ``McLean'' effect
\citep{McLean79}, where an increase of polarization due to scattered
light is seen across blueshifted P Cygni absorption
\citep[see][]{Vink02}, as here the features in \object{Hen 3-519} extend
\emph{beyond} the blueshifted part of the P Cygni absorption.

\begin{figure*}[htb]
\sidecaption
\includegraphics[width=12cm,clip]{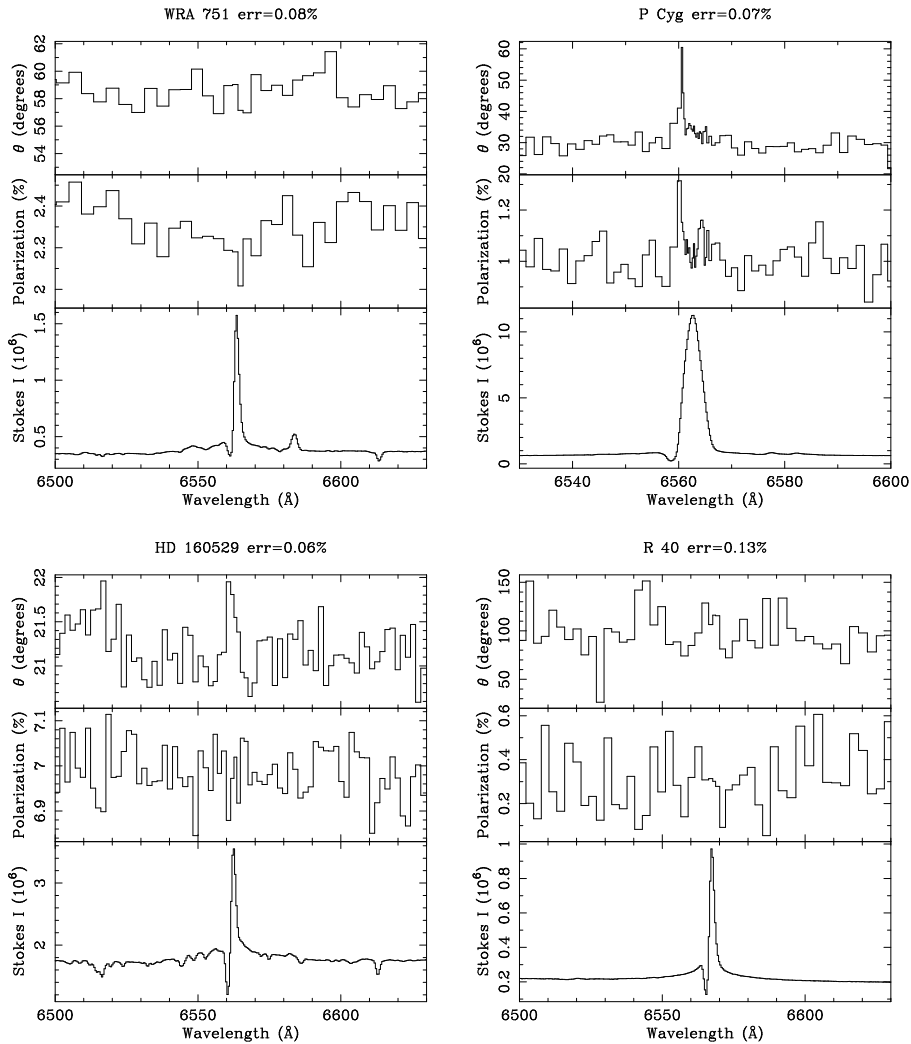}
 \caption[]{Polarization spectra of the Galactic LBVs WRA 751, \object{P Cyg}
 and \object{HD 160529}; and the SMC LBV R 40. The panels show the same as
 figure \ref{fig:etacar}, and the data has been rebinned to the
 1$\sigma$ errors stated above each plot. Note the different
 wavelength range of \object{P Cyg}. The jump in polarization seen in the
 H$\alpha$ absorption component of \object{P Cyg} is attributed to the steep
 increase in Stokes $I$ being unresolved, as explained in section
 \ref{sec:obs}. \object{HD 160529} did not display a line-effect at either
 epoch it was observed, and only the Feb 2003 data is shown here.}
\label{fig:no_le}
\end{figure*}

\paragraph{WRA 751} 
This star was first considered to be an LBV by \cite{Hu90}, but is
still referred to as an LBV candidate by many authors. It is
classified as a strong-amplitude variable by VG01, but the AAVSO
light-curve shows that it has not varied by more than 0.5mags over the
last $\sim$10 years. Radio images by \cite{D-W02} show a roughly
circular, clumpy nebula of diameter $\sim$ 20\arcsec with evidence for
a central, edge-on nebular ring of diameter $\sim$ 5\arcsec. The
object's polarization spectrum shows no line-effect (see Figure
\ref{fig:no_le}).

If the light from the star \emph{is} intrinsically polarized, the
upper limit we can put on this is rather high due to the low S/N of
the spectrum and the relative weakness of the H$\alpha$ line (see
Sect. \ref{sec:specpol}, Eq. (\ref{eq:pint})~). The sensitivity
limit for detection of dilution of the continuum by the emission line
is 0.44\%. As we have positive detections in this study of less than
0.3\%, we cannot rule out a similar level of asphericity in WRA 751's
wind.

\paragraph{P Cyg} 
Since undergoing two huge eruptions in the 17th century, there is no
evidence this star has undergone any short-timescale variability,
although \citet{dG-L92} argue the star has gradually brightened by
about 0.5 mags since then as it evolves redward. The star's nebula has
been the subject of much investigation over the last $\sim$ 10
years. It consists of an inner- and outer-shell, of diameters
22\arcsec ~and 1.6\arcmin ~respectively, both with roughly spherical
morphologies \citep{Barlow94,Meaburn96}. Optical images of the
circumstellar material reveal a complex clumpy morphology down to the
inner 0.2\arcsec \citep{Meaburn00,Chesneau00}. Multiple interferometric
radio observations have shown that the clumpiness of the inner
0.4\arcsec ~of the nebula is highly variable on timescales of days
\citep{Exter02}.

Both the observations of this object, taken 3 days apart, show changes
in polarization just blueward of the centre of the emission line, with
varying behaviour from set to set. They line up not with the P Cygni
absorption component but with the steep rise in flux from the
absorption component to the centre of the emission line, which may be
at the limit of the instrument's spectral resolution. As polarization
in this part of the spectrum varies wildly from set to set, whilst the
continuum level remains constant, we therefore explain these features
as artifacts introduced by the slightly different spectral resolutions
of the $O$ and $E$ beams emerging from the calcite. In terms of
difference in polarization between the continuum and the line-centre,
there is no change within the uncertainties (see Fig.
\ref{fig:no_le}). The detection limit is 0.15\% and 0.18\% for the
10/12/03 and 13/12/03 observations respectively. The H$\alpha$
line-centre polarization we measured agrees well with the line-centre
polarization measured by \citet{Taylor91} and \citet{Nordsieck01},
supporting the hypothesis that the H$\alpha$ line is intrinsically
unpolarized. We find no change in continuum polarization for our two
datasets taken 3 days apart.

Multiple spectropolarimetric measurements of \object{P Cyg} are described by
both Taylor et al. (20 measurements) and Nordsieck et al. (15
measurements). Both found that while the line-centre polarization
remained constant the continuum polarization was variable on
timescales of days. Both interpreted this as variable intrinsic
polarization of the object coupled with the ISP, with an intrinsically
unpolarized H$\alpha$ line as described above for \object{AG Car} and HR
Car. Taylor et al. found that the intrinsic polarization varied from
0.04\% to 0.48\% at random PAs. Such behaviour was ascribed by
Nordsieck et al. to electron scattering by a clumpy wind, with a
spherical distribution of clumps and a clump--ambient material density
contrast of $\ga$ 20. It may be the case that at the epoch of our
observations any intrinsic polarization due to wind clumps was too
small to be detected.

\paragraph{HD 160529} 
Despite being established as spectroscopically and photometrically
variable for over 30 years \citep{Wolf74}, \citet{Serk75} noted no
polarimetric variability, and it is listed by \citet{H-B82} as a
`polarized standard'. This star represents the lower-luminosity end of
the LBV instability strip and is not strongly variable, with
photometric variations of only a few tenths of a magnitude over the
last 20 years \citep{Sterken91,Stahl03}. The star has no associated
nebulosity, no nebulous emission in its spectrum and no evidence of
circumstellar dust \citep{Hutsemekers97,N-C97}. We note no [N
\textsc{ii}] emission in our spectra.

The relative weakness of the H$\alpha$ emission makes the detection of
a line effect difficult for this object (Fig. \ref{fig:no_le}. The
detection limit for a line-effect for this observation is 0.4\%.

\paragraph{R 40} 
This star is the only known LBV in the SMC. \citet{Sterken98} show
that R 40 increased steadily in brightness by 0.7 mags in the years
1985 -- 1995. AAVSO records show this object has been constant in
luminosity since $\sim$ 1995, although there is a gap in their records
of $\sim$ 3 years \citep{Waagen05}. There is presently no evidence for
associated nebular or dust emission, and we find no [N \textsc{ii}]
emission in its spectrum.

No line-effect was observed on either of the observing dates. As the
object was observed on consecutive nights and the data for each night
looked the same, the data were combined to improve the S/N, but still
no line-effect was observed (Fig. \ref{fig:no_le}. The detection limit
is 0.37\%.


\begin{figure*}[htb]
\includegraphics[width=17cm]{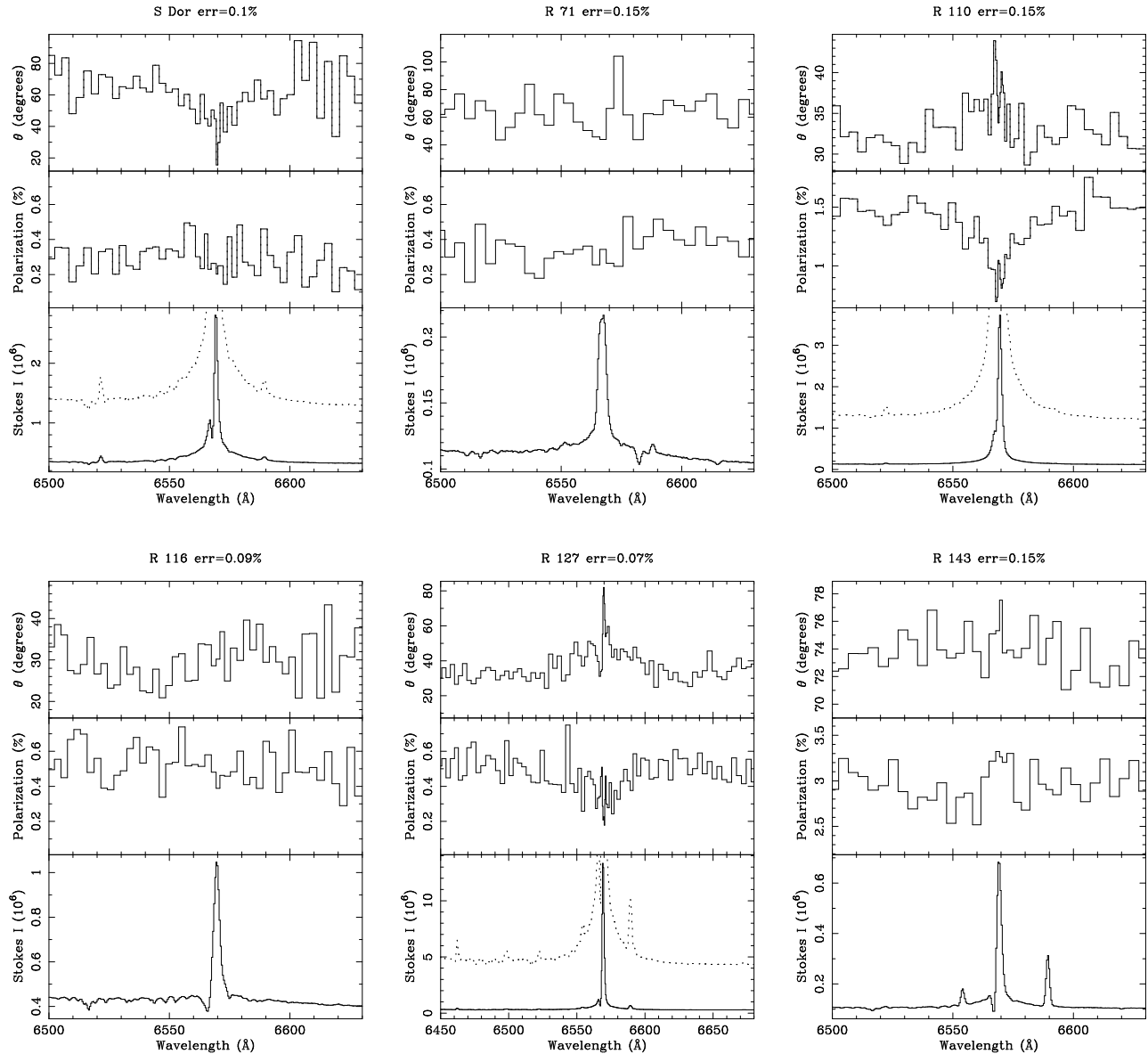}
 \caption[]{Polarization spectra of the LMC LBVs \object{S Dor}, R 71, R 110, R
 116, R 127 and R 143. The panels show the same as figure grrrr. Note
 the different wavelength scale for R 127. Where line-effects are
 observed, the Stokes $I$ spectrum is magnified to show the broad
 change in polarization associated with the wings of the emission
 line. A narrow P Cygni absorption component in the H$\alpha$ line of
 R 127 was found to introduce artifacts in the polarization spectrum
 (see Section \ref{sec:obs}), and has been edited out.}
\label{fig:lmc}
\end{figure*}

\paragraph{S Dor} 
Despite being one of the `archetypal' LBVs for many years, the star
has only recently been observed with the spectral appearance of an
F-type supergiant -- the common spectral-type of LBVs at visual
maximum \citep{Massey00}. Also, it has no directly detectable LBV
nebula. \citet{Weis03} found that the dynamical age of the large-scale
elliptical nebula surrounding \object{S Dor} meant that it was too old
to have been formed in the LBV phase; and despite broadened [N
\textsc{ii}] emission in the star's spectrum (also noted in our
spectrum), any circumstellar nebula from which it originates would
have to be less than the detection limit of $\sim$ 0.25pc in
diameter. The star's spectrum indicates a $T_{\rm eff}$ at observation
of $\sim$ 9000K, putting the star close to visual maximum. The
bolometric correction (BC) method of determining the star's
temperature is unreliable, due to disagreement over the star's
temperature and luminosity at visual minimum, and also that the star
may vary in bolometric luminosity during outbursts (see VG01). The
AAVSO light-curve of \object{S Dor} shows it to have been exhibiting
strong-amplitude variability over the last 10-15 years.

The polarization spectrum (Fig. \ref{fig:lmc}) has a suggestion of a
broad PA rotation across the line from $\sim$ 60\deg \space to $\sim$
40\deg. This corresponds to a 2$\sigma$ detection and is therefore a
borderline case. The single pixel dip in PA at the centre of the line
is not trusted as `real' behaviour, as sharp spectral features can
produce erratic data for the reasons described in Sect.
\ref{sec:obs}. The detection limit in this data is 0.2\%.

\paragraph{R 71} 
The star is noted by \citet{Weis03} as having no visually detectable
nebula and no nebular emission in its spectrum, although
\citet{Voors99} did find evidence for a warm dust component. Also, our
spectrum does show very weak [N \textsc{ii}] $\lambda$6584 emission
which suggests a nebular component. VG01 states that this star has
never been observed as cool as the classical LBV minimum temperature
of $\sim$ 8500K. AAVSO records show this star has been constant in
brightness to within $\sim$ 0.2 mags for the last $\sim$ 15 years,
although the star did increase in brightness by 0.5mags for about a
year in 2003 \citep{Waagen05}.

This object shows no line-effect (Fig. \ref{fig:lmc}). The comparitive
weakness of the H$\alpha$ line in this data means that the detection
limit is 0.4\%.

\paragraph{R 110} 
It is not known whether this star has associated circumstellar dust or
gas as it is poorly-studied. We do find, however, weak [N \textsc{ii}]
$\lambda$6584 emission in our spectrum, which suggests that the star
may have some form of nebula. The star was observed to increase in
brightness by 0.9 mags to $m_{\rm V} = 9.7$ between 1960 and 1993 by
\citet{vG97}. AAVSO photometric data on this object is limited, but it
has been seen to decrease in brightness to $m_{\rm V} \sim 11$ in the
last $\sim$ 10 years \citep{Waagen05}.

The data exhibits a broad depolarization of $\ga$ 0.5\% accompanied by
a broad PA rotation (Fig. \ref{fig:lmc}). The width of this feature
corresponds to the width of the broad wings of the emission line. The
sharp features observed in the polarization spectrum at the line
centre are spread over $\sim$ 10 pixels, which imply that they are
resolved and are judged to be real. The line-to-continuum vector in
$Q-U$ space yields an intrinsic polarization of (0.55 $\pm$ 0.09\%) at
a PA of (26 $\pm$ 5)\deg.

\paragraph{R 116} 
The similarities between this star and $\zeta_{1}$ Sco noted by
\citet{vG-S99} led them to suggest that R 116 is an
ex/dormant-LBV. AAVSO records show that the star has been constant in
brightness over the last 15 years to within 0.2 mags \citep{Waagen05},
and photometry collated from various sources by van Genderen \&
Sterken suggested the star has been constant for over 40 years. There
is presently no evidence for associated nebular or dust emission.

This object shows no line-effect (Fig. \ref{fig:lmc}). The detection
limit is 0.28\%.

\paragraph{R 127} 
The LBV nature of this star was discovered by \citet{Stahl83}. Since
then it has been observed to cool to the classical LBV minimum
temperature of $\sim$ 8000K around 1993, and is now apparently on its
way back to visual minimum \citep[VG01 \& references
therein;][]{Waagen05}. Coronagraphic imaging by \citet{Clampin93}
revealed an axisymmetric circumstellar nebula, with density
enhancements either side of the star at a PA of $\sim$ 100\deg ~and an
opening angle of 90\deg. HST observations by \citet{Weis03} reveal
evidence for asymmetric expansion within the inner nebula.

Spectropolarimetry by \citet{S-L93} revealed a high degree ($\sim$
1\%) of intrinsic, variable polarization; the PA of which remained
roughly perpendicular to the density enhancements. They explained the
roughly constant intrinsic PA and observed P Cygni profiles as
electron-scattering off an expanding equatorially-enhanced wind
aligned with the outer nebula. The variable degree of polarization and
slightly varying PA were explained by a clumpiness of the wind.

Our data shows a broad rotation in PA which corresponds to the broad
wings of the emission line (Fig. \ref{fig:lmc}). The emission line
has a narrow P Cygni absorption component which was found to introduce
erratic behaviour in the polarization spectrum. For the sake of
clarity this feature was edited out. The sharp increase in PA at the
line-centre is spread over $\sim$ 5 pixels and so are judged to be
real. The line-centre polarization has been attenuated slightly by
editing out the absorption component, but the $Q-U$ excursion
approaches the H$\alpha$ line-centre polarization measured by
\citet{S-L93}.

Taking this to be the ISP towards R 127, the intrinsic polarization is
0.50 $\pm$ 0.04\% at a PA of 20 $\pm$ 2\deg. This is in good agreement
with Schulte-Ladbeck et al. who found a PA of 24.4\deg ~in the $V$
band, although at a higher degree of polarization of 1.37\%. These
results, along with other polarization measurements collated by
Schulte-Ladbeck et al., show that the intrinsic polarization of R 127
appears to be at a roughly consistent PA of $25 \pm 10 \deg$. This
endorses their proposed model of an equatorially-enhanced, clumpy wind
viewed at a high inclination angle.

\paragraph{R 143} 
\citet{Weis03} reported that the star is located in a small,
triangular-shaped nebula of linear size $\sim$ 1.2pc within the 30 Dor
region. The disrupted shape of the nebula was speculated by Weis to be
due to strong stellar winds, high density and turbulent motions
present in 30 Dor. When observed by \citet{Nota95}, the nebula was
found to have filamentary structures eminating from the central
source, reminiscent of \object{HR Car}. However, \citet{Smith98} found that the
dynamics of these structures suggested they were not associated with
the star but of the background 30 Dor H \textsc{ii} region. They did
find evidence for a small clump of ejected material within 2\arcsec
~of the star but could not draw any conclusions as to its morphology.

This object shows no line-effect (Fig. \ref{fig:lmc}). The detection
limit is 0.34\% due to poor S/N.

\section{Discussion \label{sec:disc}}

\begin{figure}[ht]
\resizebox{\hsize}{!}{\includegraphics[angle=-90,bb=0 20 513 725,clip]{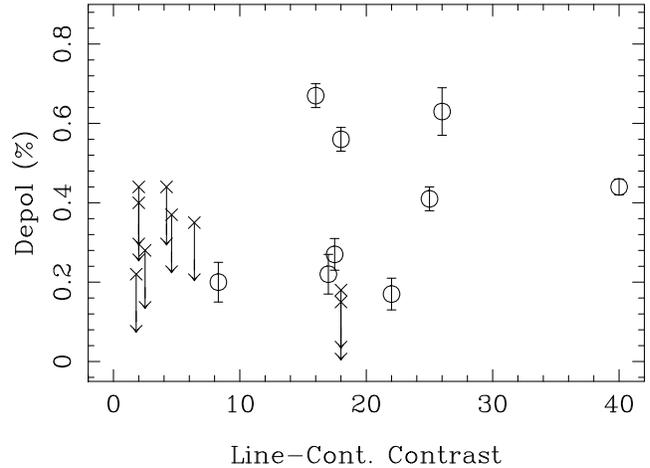}}
 \caption[]{Strength of depolarization against the line-to-continuum
 contrast of the H$\alpha$ line. Circles show line-effects, crosses
 show non-detections. Where no depolarization was observed, the upper
 limit for an undetected line-effect is shown. For the sake of
 clarity, \object{$\eta$ Car} is not shown as its depolarization and
 line contrast are both much greater than any other object.}
\label{fig:pol_rel1}
\end{figure}

Of the 11 confirmed LBVs, we have 5 positive detections of
line-effects plus the borderline case of \object{S Dor}.  Of the three LBV
candidates, only \object{Hen 3-519} shows a change in polarization across its
emission lines, and the behaviour of this star's polarization spectrum
cannot simply be explained by the dilution of the polarized flux by
the emission line. Of the non-detection objects, in some cases the
detection limits exceed the levels of depolarization observed in
objects $with$ line-effects.  There may therefore be similar levels of
depolarization in some of these objects that have gone undetected.
Even if all LBVs had aspherical winds, we would expect some
non-detections due to inclination and cancellation effects
e.g. face-on disk, or two clumps at 90\deg ~to each other.  We
therefore put the rate of wind-asphericity in LBVs at $\ga$ 50\%,
higher than that of the other classes of evolved massive star that
have been studied comprehensively, O supergiants and Wolf-Rayet stars
{\citep[25\% and 20\% of sample sizes 20 and 16
respectively,][]{Harries02,Harries98}.

Rotation and mass loss play intricate roles in the evolution of
massive stars \citep{Langer97,M-M00}. Given the transitional role of
LBVs within the evolutionary paths of massive stars, the high
incidence of intrinsic polarizations that we find in comparison to
those of O/WR stars is expected to reveal telling information about
the onset of wind asymmetries. To probe the origin of the wind
asymmetries, we correlate our detections with a range of LBV
parameters (Sect. 4.1), before we discuss possible scenarios in
Sect. 4.2.

\subsection{Correlations}

In terms of temperature and phase, no correlation is observed between
detections and non-detections -- line-effects are observed at visual
minimum (\object{HR Car}), maximum (R 110), and intermediate (e.g. R 127)
phases. No correlation is observed with luminosity, although we point
out that the luminosities of these objects (even those in the Clouds)
are uncertain, with different authors stating different values for L/
$\lsun$. VG01 states that luminosities of Galactic objects are
uncertain to 0.2-0.3 dex, while MC objects are uncertain to $\sim$ 0.1
dex.

\subsubsection{Line-strength -- wind asphericity relation?}

Inspection of Table \ref{tab:Hadata} and Figure \ref{fig:pol_rel1}
shows that the objects showing line-effects all have strong H$\alpha$
lines. Indeed, the strongest H$\alpha$ line definitely showing no
line-effect is R 143, with a line-to-continuum contrast of 6.4. \object{S Dor}
has a line contrast of 8.3, and has a borderline case for a
line-effect. Only \object{P Cyg} of the strong-emission objects has no
clear-cut line-effect. From this we could draw one of two conclusions:
either the method is not sensitive enough to detect line-effects in
weak emission lines; or objects with strong H$\alpha$ lines have
strong asphericity in their winds. Line-effects (albeit weak) are seen
across the He \textsc{i} $\lambda$6678 line in the spectrum of \object{AG Car}
(contrast $\sim$ 1.7), the weaker lines in }'s spectrum
($\sim$ 1.7), and the weak Fe \textsc{ii} lines in the spectrum of HR
Car ($\sim$ 1.5). This therefore leads us to discount the first
explanation and suggest that LBV winds producing stronger emission are
more likely to have an aspherical geometry.

\begin{figure}[ht]
\resizebox{\hsize}{!}{\includegraphics[angle=-90,bb=0 20 513 725,clip]{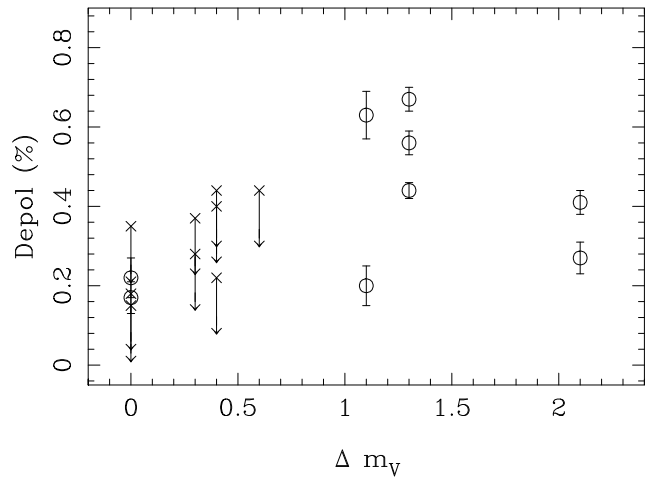}}
 \caption[]{Strength of depolarization against apparent variability
 over the last ten years. The symbols show the same as Figure
 \ref{fig:pol_rel1}.}
\label{fig:pol_rel2}
\end{figure}

\subsubsection{Galactic vs. Extra-Galactic LBVs}

Of the Galactic objects observed, 4 out of 7 show line-effects. Of the
three non-detections, one (\object{P Cyg}) has been shown to exhibit a
change in polarization across its emission lines at earlier epochs
\citep{Taylor91,Nordsieck01}. By comparison, only 2 out of 7
Magellanic Cloud objects show line-effects, with one borderline case
(\object{S Dor}). One explanation for this difference may be that the
selection of Galactic LBVs is not truly representative of the LBV
phenomenon. \object{Hen 3-519} is commonly thought-of as a post-LBV
object, indeed its broad emission lines are reminiscent of a
Wolf-Rayet star. WRA 751 is located in a distinctly separate location
on the H-R diagram to the other LBVs (see Figure \ref{fig:hrdiag}) and
has not been observed to show variability on the same scales as other
LBVs such as \object{AG Car} or \object{S Dor}
\citep{vG92}. \object{$\eta$ Car} and \object{P Cyg} are unusual
objects even by LBV standards -- whilst both had huge outbursts
hundreds of years ago, \object{P Cyg} has been roughly constant in
luminosity for many years, while \object{$\eta$ Car}'s variability is
often explained by the expanding nebula moving across our line of
sight, coupled with the effects of being in a highly eccentric binary
system \citep{I-dG99,D-H97}. This leaves \object{AG Car}, \object{HR
Car} and \object{HD 160529} as the only `normal' Galactic LBVs. We may
expect the MC LBVs to be a more homogeneous group, but as the data
quality for the MC LBVs is not as good as for the Galactic ones, and
because of the low sample numbers involved, we draw no conclusions as
to differences between these two populations at the current stage.

\subsubsection{Wind asphericity and strong-amplitude variability}

\begin{figure*}[tb]
\sidecaption
\includegraphics[bb=10 10 541 390,width=12cm,angle=0,clip]{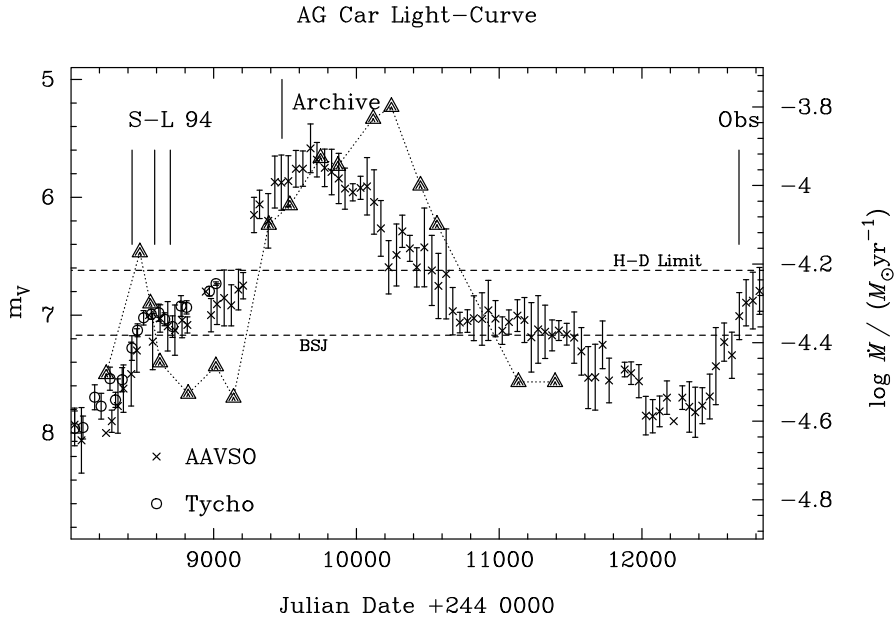}
 \caption[]{The light curve of \object{AG Car} for 1989 -- 2004. Circles show
data from the Tycho mission, crosses show validated data from the
AAVSO records. The data has been rebinned to 50 days, and the error
bars show the 1$\sigma$ standard deviations. The dates of our
observation, the archive observation, plus the observations of SL94
have been marked. Also marked are the brightnesses at which the star
is crossing the empirical H-D limit, and the bi-stability jump (BSJ)
at $\sim$ 21,000K (under the assumption that the star varies at
constant bolometric luminosity). Also marked on the plot are the
mass-loss rates derived by \citet{Stahl01} (triangles).  }
\label{fig:agcar_lc}
 \end{figure*}

A correlation can be found by looking at recent light-curves of the
objects. Observations from the last $\sim$ 15 years obtained from the
AAVSO and the $Tycho$ mission show that, of the objects showing
definite line-effects, all show variability of $\ga$ 1 mag (see Fig.
\ref{fig:pol_rel2}). Of the others, only \object{S Dor} shows this
kind of variability, and this object was a borderline detection. We
therefore propose a link between wind asphericity and recent, strong
variability. Figure \ref{fig:agcar_lc} shows the light-curve of AG
Car, going back 15 years. Marked on the plot are the H-D limit and the
bi-stability jump \citep[BSJ, see Sect. \ref{sec:intro} and
][]{Lamers95,Vink99}, under the assumption that the star varies at
constant bolometric luminosity. It can be seen that \object{AG Car}
has crossed its H-D limit and spent long periods very close to the BSJ
during this time. The same is true for R 127 \citep{Waagen05} and
\object{S Dor}, although the assumption of constant bolometric
luminosity may break down for S Dor (VG01). Whilst the luminosities of
\object{HR Car} and R 110 mean they are below the H-D limit (leaving
the uncertainties in their luminosities aside), both stars have shown
recent strong variability \citep[AAVSO,][]{Waagen05}, and are close to
a second BSJ proposed by \citet{Lamers95} at $\sim$ 10,000K, as Fe
\textsc{iii} recombines to Fe \textsc{ii}. Proximity to the BSJs
coupled with a variable effective temperature may play a part in
producing wind asphericity.

\subsection{Interpretation of polarimetric variability}

We have shown that at least half of the LBVs observed display evidence
for intrinsic polarization. Of the 4 objects for which multi-epoch
observations exist, all are polarimetrically variable, with both the
degree of polarization and PA changing on timescales of $\sim$ a week
to months. As the bulk of the polarization occurs within a couple of
stellar radii, variable wind properties of an outflow moving at a few
hundred \kms ~could produce variability on these timescales. Here we
will discuss four different possible scenarios for producing such
behaviour.

\subsubsection{Flip-flopping wind}

In this scenario, the wind flips from an equatorially-enhanced flow to
a bi-polar flow or vice versa. Such a scenario could occur if the star
was close to a BSJ and was rotating close to break-up, producing a
temperature differential between the equator and the poles. As the
star's global $T_{\rm eff}$ changes, one of these regions may reach
the BSJ temperature before the other, changing the geometry of the
wind. As the plane of the equator and poles are perpendicular to each
other, at this point we may observe a 90\deg ~flip in intrinsic
polarization. At intermediate stages we would expect the polarization
of the two components to cancel each other out, and so we would
observe lower levels of polarization but at PAs more or less aligned
with one of the two axes.

The results of SL94 showed that, for three separate observations of AG
Car over $\sim$ 2 years, the PA of the intrinsic polarization flipped
roughly by 90\deg ~from 55\deg ~to 133\deg ~and back again. From
Fig. \ref{fig:agcar_lc} we can see that the star brightened by $\sim$
0.2mags between their first and second observation, then had dimmed by
$\sim$ 0.1mags by the time of their third. We can also see that the
star was close to its BSJ at this time, assuming the star varies at
constant luminosity. This alone is consistent with a flip-flopping
wind. However, when one looks at the PAs measured at 1994 (26\deg) and
2003 (84\deg), as well as those measured by \citet{Leitherer94}, these
angles are completely unrelated; whilst the degree of polarization
remains in the range 0.4-0.6\%. This suggests that the results of SL94
may have been a result of limited temporal sampling, and that the
flip-flopping wind cannot be the sole explanation of the polarization
behaviour. A similar situation is observed for \object{HR Car} and
\object{P Cyg}, where the PA appears random with time and bears no
relation to the level of polarization.

\subsubsection{Opacity changes within the wind}

As explained in Sect. \ref{sec:intro}, single scattering off a
flattened wind would produce polarization perpendicular to the plane
of scattering. However, if the opacity was to increase, multiple
scattering effects would come into play. It has been shown by
e.g. \citet{Angel69} and \citet{Wood96} that multiple scattering in an
optically thick, flattened wind can produce polarization
\emph{parallel} to the plane of scattering. If the density of the
inner regions of the wind were to increase due to e.g. an increase in
mass-loss rate, we may expect to see the PA of polarization change
from perpendicular to parallel to the flattened wind's major axis. As
with the flip-flopping wind, we would expect intermediate stages to
have low degrees of polarization in one of the two planes due to
cancellation effects.

In Figure \ref{fig:agcar_lc}, the mass-loss rate for \object{AG Car}
as determined by \citet{Stahl01} are overplotted on the star's light
curve. We can see that between the first two measurements by SL94, the
mass-loss rate drops, coinciding with a 90\deg ~flip in the star's
intrinsic polarization. However, at the time of SL94's third
observation, at which point the polarization had flipped back, the
mass-loss rate had remained roughly constant. The seemingly unrelated
PAs at similar levels of polarization of the later observations are
further reason to disregard this scenario for \object{AG Car}. The
random temporal behaviour of \object{HR Car} and \object{P Cyg} also
cannot be explained this way.

\subsubsection{Clumpy wind}

Here, the wind is not a smooth homogeneous outflow but instead
consists of localized density enhancements, or `clumps'. The bulk of
the scattering would be due to the closest, densest clumps. If the
clumps were ejected in a spherical distribution, or in a flattened
wind which was observed at a low inclination, the PA of the intrinsic
polarization would be related to the projected angle between the star
and the clump, and appear effectively random with time. Only in a
flattened wind observed at a high inclination angle would the PA
remain roughly constant. Multiple clumps within the wind could serve
to increase the intrinsic polarization or cancel each other out,
making the degree of polarization also seemingly random with
time. Wind-clumping has previously been invoked to explain the random
polarimetric variations observed in Wolf-Rayet (WR) stars
\citep{Robert89}.

\object{AG Car}, \object{HR Car}, and \object{P Cyg} are all described
in this paper to have intrinsic polarization which appears to be
essentially random with time, both in level and PA. This is consistent
with the clumpy wind scenario, with the clumps ejected in a roughly
spherical distribution on the plane of the sky. Only R 127 has
polarization with PA which remains roughly constant at 25 $\pm$
10\deg. This is consistent with a clumpy wind ejected on a
significantly flattened plane on the sky. The P Cygni profiles
observed in the star's spectrum imply that material is being ejected
towards us along the line-of-sight, therefore the favoured explanation
of \citet{S-L93} was that we are looking at an edge-on,
equatorially-enhanced, clumpy wind. Our latest measurement supports
this model.

The strength of the polarimetric variability of LBVs is greater than
  that observed in WRs \citep{Robert89}. It is interesting to note
  that Robert et al. find that the amplitude of the variability in WRs
  increases with {\em decreasing} terminal wind speed. This is
  consistent with the greater variability of LBVs ($\sim$ 1\%,
  $v_{\infty} \sim 200$\kms) compared with WRs ($\la$ 0.2\%,
  $v_{\infty} \ga 1000$\kms).

Investigations into the physical parameters of a single clump required
to produce the observed levels of polarization can be found in
\citet{Nordsieck01} and \citet{R-M00}. The solid angle subtended by
the clump must be large enough to scatter a significant fraction of
the starlight. \citet{Nordsieck01} find that the clump must be within
$\sim$ 2.5$R_{\rm star}$ and be $\sim$ 20 times as dense as the
ambient wind; while \citet{R-M00} show that the clump must be a
significant fraction of the size of the star \nolinebreak ($R_{\rm
clump} \ga 0.5 R_{\rm star}$). Modelling the observed polarisation
variablility of an LBV wind with multiple clumps at various distances
in the wind and under a range of geometric distributions (inclination
and opening angle) would yield valuable information about the overall
wind geometry, as well as the clump production rate.

\subsubsection{Binarity} 

If the variability of the intrinsic polarization were to be periodic,
this would suggest that the polarimetric variations were due to
interactions between the LBV and a binary companion, {\`a} la close
WR-O systems. Here, the polarization is produced by the electrons in
the WR star's fast, ionised wind scattering the light from the bright
companion O star in the region between the two. The polarization
describes a double loop in $Q-U$ space about a central locus for each
orbital cycle \citep[see][]{StL87}. The stars must be sufficiently
close together in order for the WR's wind to intercept enough of the O
star's light, and consequently the systems have periods of a few days
to months. The temporal resolution of present polarimetric
observations is insufficient to detect such oscillations in the LBVs'
polarization, and so this scenario warrants further invesigation. If a
similar mechanism is responsible for the LBV's polarization, the LBV
would either have to play the role of the scatterer (i.e. be in a
system with a hot star) or the illuminator (i.e. have a WR companion).

If the LBVs were in a system with a WR star, we may expect to see i)
X-rays from the collision between the fast WR wind and the LBV wind,
cf. the WC8 -- evolved O star system \object{HD 68273}
\citep{Schild04}; and ii) other periodic phenomena such as
spectroscopic oscillations or the cyclical photometric variability due
to the comparable mass of the companion. We note that of the four
polarimetrically variable LBVs, only \object{P Cyg} has been detected
as an X-ray source \citep{Berghofer00}, and this has been rated as
questionable by the authors. These stars have been subject to
time-resolved photometric and spectroscopic monitoring, but while
oscillations have been detected, these are on characteristic
timescales rather than strictly periodic, and are explained as stellar
pulsations \citep[see e.g.][]{Stahl01,Lamers98}.

If the LBV's wind was electron-scattering the light of a companion, the
luminosity of the companion would have to be comparable to that of the
LBV in order for the scattered light to make up the $\sim$ 1\%
polarization of the overall light we observe (the scattering of an O
star's light in WR-O binaries results in typically $\Delta P \sim
0.4\%$). It is unlikely that in this case the companion would not have
been already detected.

As an aside, it has been shown that after the periodic components have
been subtracted, the polarimetric variations in WR-O binary systems
are no greater than for single WR stars \citep{Robert89}. As these
polarimetric variations are associated with wind-clumping in WR stars,
this result implies that binarity does not play a significant role in
the clumping of hot star winds. In short, binarity is an unlikely
explanation for the observed polarimetric variability, either as a
scattering mechanism or as a catalyst for wind-clumping.

\section{Conclusions \label{sec:conc}}

From our observations, it is clear that LBVs show a high rate of
wind-asphericity. We find evidence for intrinsic polarization in $\ga$
50\% of those stars studied, compared with 25\% and 20\% for similar
studies of O supergiants and Wolf-Rayet stars respectively. However,
multiple observations reveal levels and angles of intrinsic
polarization that are seemingly random with time. In investigating
four possible explanations for this we argue the evidence points away
from the simple explanation of equatorially-enhanced or bi-polar
flows. Instead, we interpret the variable intrinsic polarization as
evidence for significant clumping within the wind. The random levels
of polarization measured in R 127 are confined to position angles
(PAs) of $\pm$ 10\deg, which can be explained as an equatorially
enhanced, clumpy wind viewed edge-on. If viewed at lower inclination
angles, the measured PAs may seem as random as those observed in AG
Car, \object{HR Car} and \object{P Cyg} at limited temporal
sampling. Therefore, time-resolved spectropolarimetric monitoring of
LBVs would be required in order to detect axi-symmetry in their winds.

We find that wind asphericity is more likely to be found in stars with
strong H$\alpha$ emission and strong recent variability ($\ga$
1mag). As both these properties are linked to the stars' mass-loss
rates they may also be linked to the clumping of the wind. With higher
mass-loss rates, slower winds and lower effective gravities than their
evolutionary neighbours, as well as residing in the bi-stability
temperature regime, small changes in radius and/or effective
temperature may lead to significant wind instabilities and
inhomogeneities. This is consistent with the lower incidence of
intrinsic polarization found in O supergiants and Wolf-Rayet stars.


\begin{acknowledgements}

We thank the anonymous referee and Norbert Langer for their helpful
comments and suggestions.  We wish to thank the staff at the AAT and
WHT for their assistance during the observing runs. This work has made
extensive use of the online database of \hbox{\textsc{simbad}} and the
Starlink software suite, in particular the packages \textsc{figaro},
\textsc{tsp} and \textsc{polmap}. We have made use of data from the
AAT archive and the $\eta$ Car \emph{HST} Treasury Project. We
acknowledge with thanks the variable star observations from the AAVSO
International Database contributed by observers worldwide and used in
this research. BD and JSV are funded by PPARC.
      
\end{acknowledgements}

\bibliographystyle{aa}
\bibliography{aa_2005_2781}

\begin{thebibliography}{86}
\expandafter\ifx\csname natexlab\endcsname\relax\def\natexlab#1{#1}\fi

\bibitem[{Aitken \& Hough(2001)}]{A-H01}
Aitken, D.~K. \& Hough, J.~H. 2001, PASP, 113, 1300

\bibitem[{{Angel}(1969)}]{Angel69}
{Angel}, J.~R.~P. 1969, \apj, 158, 219

\bibitem[{{Barlow} {et~al.}(1994){Barlow}, {Drew}, {Meaburn}, \&
  {Massey}}]{Barlow94}
{Barlow}, M.~J., {Drew}, J.~E., {Meaburn}, J., \& {Massey}, R.~M. 1994, \mnras,
  268, L29+

\bibitem[{{Bergh{\" o}fer} \& {Wendker}(2000)}]{Berghofer00}
{Bergh{\" o}fer}, T.~W. \& {Wendker}, H.~J. 2000, Astronomische Nachrichten,
  321, 249

\bibitem[{Cassinelli {et~al.}(1987)Cassinelli, Nordsieck, \&
  Murison}]{Cassinelli87}
Cassinelli, J.~P., Nordsieck, K.~H., \& Murison, M.~A. 1987, ApJ, 317, 290

\bibitem[{{Chesneau} {et~al.}(2000){Chesneau}, {Roche}, {Boccaletti}, {Abe},
  {Moutou}, {Charbonnier}, {Aime}, {Lant{\' e}ri}, \& {Vakili}}]{Chesneau00}
{Chesneau}, O., {Roche}, M., {Boccaletti}, A., {et~al.} 2000, \aaps, 144, 523

\bibitem[{Clampin {et~al.}(1993)Clampin, Nota, Golimowski, Leitherer, \&
  Durrance}]{Clampin93}
Clampin, M., Nota, A., Golimowski, D.~A., Leitherer, C., \& Durrance, S.~T.
  1993, \apj, 410, 35

\bibitem[{Clampin {et~al.}(1995)Clampin, Schulte-Ladbeck, Nota, Robberto,
  Paresce, \& Clayton}]{Clampin95}
Clampin, M., Schulte-Ladbeck, R.~E., Nota, A., {et~al.} 1995, AJ, 110, 251

\bibitem[{{Cox} {et~al.}(1995){Cox}, {Mezger}, {Sievers}, {Najarro},
  {Bronfman}, {Kreysa}, \& {Haslam}}]{Cox95}
{Cox}, P., {Mezger}, P.~G., {Sievers}, A., {et~al.} 1995, \aap, 297, 168

\bibitem[{Damineli {et~al.}(2000)Damineli, Kaufer, Wolf, Stahl, Lopes, \&
  de~Araújo}]{Damineli00}
Damineli, A., Kaufer, A., Wolf, B., {et~al.} 2000, \apjl, 528, 101

\bibitem[{Davidson \& Humphreys(1997)}]{D-H97}
Davidson, K. \& Humphreys, R.~M. 1997, ARA\&A, 35, 1

\bibitem[{Davidson {et~al.}(1993)Davidson, Humphreys, Hajian, \&
  Terzian}]{Davidson93}
Davidson, K., Humphreys, R.~M., Hajian, A., \& Terzian, Y. 1993, \apj, 411, 336

\bibitem[{{de Groot} \& {Lamers}(1992)}]{dG-L92}
{de Groot}, M. \& {Lamers}, H.~J.~G.~L.~M. 1992, Lecture Notes in Physics,
  Berlin Springer Verlag, 401, 121

\bibitem[{de~Koter {et~al.}(1996)de~Koter, Lamers, \& Schmutz}]{dK96}
de~Koter, A., Lamers, H. J. G. L.~M., \& Schmutz, W. 1996, \aap, 306, 501

\bibitem[{Duncan \& White(2002)}]{D-W02}
Duncan, R.~A. \& White, S.~M. 2002, MNRAS, 330, 63

\bibitem[{{Dwarkadas} \& {Balick}(1998)}]{D-B98}
{Dwarkadas}, V.~V. \& {Balick}, B. 1998, \aj, 116, 829

\bibitem[{Dwarkadas \& Owocki(2002)}]{D-O02}
Dwarkadas, V.~V. \& Owocki, S.~P. 2002, \apj, 581, 1337

\bibitem[{{Exter} {et~al.}(2002){Exter}, {Watson}, {Barlow}, \&
  {Davis}}]{Exter02}
{Exter}, K.~M., {Watson}, S.~K., {Barlow}, M.~J., \& {Davis}, R.~J. 2002,
  \mnras, 333, 715

\bibitem[{{Frank} {et~al.}(1995){Frank}, {Balick}, \& {Davidson}}]{Frank95}
{Frank}, A., {Balick}, B., \& {Davidson}, K. 1995, \apjl, 441, L77

\bibitem[{Harries {et~al.}(1998)Harries, Hillier, \& Howarth}]{Harries98}
Harries, T.~J., Hillier, D.~J., \& Howarth, I.~D. 1998, MNRAS, 296, 1072

\bibitem[{Harries \& Howarth(1996)}]{H-H96}
Harries, T.~J. \& Howarth, I.~D. 1996, A\&A, 310, 533

\bibitem[{Harries {et~al.}(2002)Harries, Howarth, \& Evans}]{Harries02}
Harries, T.~J., Howarth, I.~D., \& Evans, C.~J. 2002, \mnras, 337, 341

\bibitem[{Hsu \& Breger(1982)}]{H-B82}
Hsu, J.-C. \& Breger, M. 1982, \apj, 262, 732

\bibitem[{Hu {et~al.}(1990)Hu, de~Winter, Th\'{e}, \& Perez}]{Hu90}
Hu, J.~Y., de~Winter, D., Th\'{e}, P.~S., \& Perez, M.~R. 1990, \aap, 227, 17

\bibitem[{Humphreys \& Davidson(1994)}]{H-D94}
Humphreys, R.~M. \& Davidson, K. 1994, PASP, 106, 1025

\bibitem[{{Humphreys} {et~al.}(1999){Humphreys}, {Davidson}, \&
  {Smith}}]{Humphreys99}
{Humphreys}, R.~M., {Davidson}, K., \& {Smith}, N. 1999, \pasp, 111, 1124

\bibitem[{Hutsemekers(1997)}]{Hutsemekers97}
Hutsemekers, D. 1997, in Luminous Blue Variables: Massive Stars in Transition,
  ed. A.~Nota \& H.~Lamers (ASP Conference Series v.120)

\bibitem[{Ishibashi {et~al.}(2003)Ishibashi, Gull, Davidson, Smith, Lanz,
  Lindler, Feggans, Verner, Woodgate, Kimble, Bowers, Kraemer, Heap, Danks,
  Maran, Joseph, Kaiser, Linsky, Roesler, \& Weistrop}]{Ishi03}
Ishibashi, K., Gull, T.~R., Davidson, K., {et~al.} 2003, AJ, 125, 3222

\bibitem[{Israelian \& de~Groot(1999)}]{I-dG99}
Israelian, G. \& de~Groot, M. 1999, \ssr, 90, 493

\bibitem[{{Lamers} \& {Pauldrach}(1991)}]{L-P91}
{Lamers}, H.~J.~G. \& {Pauldrach}, A.~W.~A. 1991, \aap, 244, L5

\bibitem[{Lamers {et~al.}(1998)Lamers, Bastiaanse, Aerts, \& Spoon}]{Lamers98}
Lamers, H. J. G. L.~M., Bastiaanse, M.~V., Aerts, C., \& Spoon, H. W.~W. 1998,
  \aap, 335, 605

\bibitem[{Lamers \& Cassinelli(1999)}]{L-C99}
Lamers, H. J. G. L.~M. \& Cassinelli, J.~P. 1999, Introduction to Stellar Winds
  (Cambridge)

\bibitem[{Lamers {et~al.}(1995)Lamers, Snow, \& Lindholm}]{Lamers95}
Lamers, H. J. G. L.~M., Snow, T.~P., \& Lindholm, D.~M. 1995, \apj, 455, 269

\bibitem[{{Langer}(1997)}]{Langer97}
{Langer}, N. 1997, in Luminous Blue Variables: Massive Stars in Transition, ed.
  A.~{Nota} \& H.~J.~G.~L.~M. {Lamers} (ASP Conference Series v.120), 83

\bibitem[{{Langer} {et~al.}(1999){Langer}, {Garc{\'{\i}}a-Segura}, \& {Mac
  Low}}]{Langer99}
{Langer}, N., {Garc{\'{\i}}a-Segura}, G., \& {Mac Low}, M. 1999, \apjl, 520,
  L49

\bibitem[{Leitherer {et~al.}(1994)Leitherer, Allen, Altner, Damineli, Drissen,
  Idiart, Lupie, Nota, Robert, Schmutz, \& Shore}]{Leitherer94}
Leitherer, C., Allen, R., Altner, B., {et~al.} 1994, \apj, 428, 292

\bibitem[{{Maeder} \& {Meynet}(2000)}]{M-M00}
{Maeder}, A. \& {Meynet}, G. 2000, \aap, 361, 159

\bibitem[{Massey(2000)}]{Massey00}
Massey, P. 2000, \pasp, 112, 144

\bibitem[{{McLean}(1979)}]{McLean79}
{McLean}, I.~S. 1979, \mnras, 186, 265

\bibitem[{{Meaburn} {et~al.}(1996){Meaburn}, {Lopez}, {Barlow}, \&
  {Drew}}]{Meaburn96}
{Meaburn}, J., {Lopez}, J.~A., {Barlow}, M.~J., \& {Drew}, J.~E. 1996, \mnras,
  283, L69

\bibitem[{{Meaburn} {et~al.}(2000){Meaburn}, {O'connor}, {L{\' o}pez}, {Bryce},
  {Redman}, \& {Noriega-Crespo}}]{Meaburn00}
{Meaburn}, J., {O'connor}, J.~A., {L{\' o}pez}, J.~A., {et~al.} 2000, \mnras,
  318, 561

\bibitem[{{Morris} {et~al.}(1999){Morris}, {Waters}, {Barlow}, {Lim}, {de
  Koter}, {Voors}, {Cox}, {de Graauw}, {Henning}, {Hony}, {Lamers}, {Mutschke},
  \& {Trams}}]{Morris99}
{Morris}, P.~W., {Waters}, L.~B.~F.~M., {Barlow}, M.~J., {et~al.} 1999, \nat,
  402, 502

\bibitem[{{Morse} {et~al.}(1998){Morse}, {Davidson}, {Bally}, {Ebbets},
  {Balick}, \& {Frank}}]{Morse98}
{Morse}, J.~A., {Davidson}, K., {Bally}, J., {et~al.} 1998, \aj, 116, 2443

\bibitem[{Nordsieck {et~al.}(2001)Nordsieck, Wisniewski, Babler, Meade,
  Anderson, Bjorkman, Code, Fox, Johnson, Weitenbeck, \& Zellner}]{Nordsieck01}
Nordsieck, K.~H., Wisniewski, J., Babler, B.~L., {et~al.} 2001, in P Cygni
  2000: 400 Years of Progress, ed. M.~de~Groot \& C.~Sterken (ASP Conference
  Series v.233)

\bibitem[{Nota \& Clampin(1997)}]{N-C97}
Nota, A. \& Clampin, M. 1997, in Luminous Blue Variables: Massive Stars in
  Transition, ed. A.~Nota \& H.~Lamers (ASP Conference Series v.120)

\bibitem[{Nota {et~al.}(1992)Nota, Leitherer, Clampin, Greenfield, \&
  Golimowski}]{Nota92}
Nota, A., Leitherer, C., Clampin, M., Greenfield, P., \& Golimowski, D.~A.
  1992, \apj, 398, 621

\bibitem[{Nota {et~al.}(1995)Nota, Livio, Clampin, \& Schulte-Ladbeck}]{Nota95}
Nota, A., Livio, M., Clampin, M., \& Schulte-Ladbeck, R. 1995, ApJ, 448, 788

\bibitem[{Nota {et~al.}(1997)Nota, Smith, Pasquali, Clampin, \&
  Stroud}]{Nota97}
Nota, A., Smith, L., Pasquali, A., Clampin, M., \& Stroud, M. 1997, ApJ, 486,
  338

\bibitem[{{Oudmaijer} \& {Drew}(1999)}]{Oudmaijer99}
{Oudmaijer}, R.~D. \& {Drew}, J.~E. 1999, \mnras, 305, 166

\bibitem[{{Oudmaijer} {et~al.}(1998){Oudmaijer}, {Proga}, {Drew}, \& {de
  Winter}}]{Oudmaijer98}
{Oudmaijer}, R.~D., {Proga}, D., {Drew}, J.~E., \& {de Winter}, D. 1998,
  \mnras, 300, 170

\bibitem[{Parthasarathy {et~al.}(2000)Parthasarathy, Jain, \& Bhatt}]{Parth00}
Parthasarathy, M., Jain, S.~K., \& Bhatt, H.~C. 2000, \aap, 355, 221

\bibitem[{{Pelupessy} {et~al.}(2000){Pelupessy}, {Lamers}, \&
  {Vink}}]{Pelupessy00}
{Pelupessy}, I., {Lamers}, H.~J.~G.~L.~M., \& {Vink}, J.~S. 2000, \aap, 359,
  695

\bibitem[{{Robert} {et~al.}(1989){Robert}, {Moffat}, {Bastien}, {Drissen}, \&
  {St.-Louis}}]{Robert89}
{Robert}, C., {Moffat}, A.~F.~J., {Bastien}, P., {Drissen}, L., \& {St.-Louis},
  N. 1989, \apj, 347, 1034

\bibitem[{{Rodrigues} \& {Magalh{\~ a}es}(2000)}]{R-M00}
{Rodrigues}, C.~V. \& {Magalh{\~ a}es}, A.~M. 2000, \apj, 540, 412

\bibitem[{{Schild} {et~al.}(2004){Schild}, {G{\" u}del}, {Mewe}, {Schmutz},
  {Raassen}, {Audard}, {Dumm}, {van der Hucht}, {Leutenegger}, \&
  {Skinner}}]{Schild04}
{Schild}, H., {G{\" u}del}, M., {Mewe}, R., {et~al.} 2004, \aap, 422, 177

\bibitem[{Schmidt-Kaler(1982)}]{S-K82}
Schmidt-Kaler, T. 1982, Landolt-Bornstein Gp. 6, Vol. 2-2b (Springer-Verlag,
  Berlin)

\bibitem[{Schulte-Ladbeck {et~al.}(1994)Schulte-Ladbeck, Clayton, Hillier,
  Harries, \& Howarth}]{S-L94}
Schulte-Ladbeck, R.~E., Clayton, G.~C., Hillier, D.~J., Harries, T.~J., \&
  Howarth, I.~D. 1994, ApJ, 429, 846

\bibitem[{Schulte-Ladbeck {et~al.}(1993)Schulte-Ladbeck, Leitherer, Clayton,
  Robert, Meade, Drissen, Nota, \& Schmutz}]{S-L93}
Schulte-Ladbeck, R.~E., Leitherer, C., Clayton, G.~C., {et~al.} 1993, \apj,
  407, 723

\bibitem[{{Schulte-Ladbeck} {et~al.}(1999){Schulte-Ladbeck}, {Pasquali},
  {Clampin}, {Nota}, {Hillier}, \& {Lupie}}]{S-L99}
{Schulte-Ladbeck}, R.~E., {Pasquali}, A., {Clampin}, M., {et~al.} 1999, \aj,
  118, 1320

\bibitem[{Serkowski {et~al.}(1975)Serkowski, Mathewson, \& Ford}]{Serk75}
Serkowski, K., Mathewson, D.~L., \& Ford, V.~L. 1975, \apj, 196, 261

\bibitem[{{Smith} {et~al.}(1998){Smith}, {Nota}, {Pasquali}, {Leitherer},
  {Clampin}, \& {Crowther}}]{Smith98}
{Smith}, L.~J., {Nota}, A., {Pasquali}, A., {et~al.} 1998, \apj, 503, 278

\bibitem[{{Smith} {et~al.}(2003{\natexlab{a}}){Smith}, {Davidson}, {Gull},
  {Ishibashi}, \& {Hillier}}]{Smith03apj}
{Smith}, N., {Davidson}, K., {Gull}, T.~R., {Ishibashi}, K., \& {Hillier},
  D.~J. 2003{\natexlab{a}}, \apj, 586, 432

\bibitem[{{Smith} {et~al.}(2003{\natexlab{b}}){Smith}, {Gehrz}, {Hinz},
  {Hoffmann}, {Hora}, {Mamajek}, \& {Meyer}}]{Smith03aj}
{Smith}, N., {Gehrz}, R.~D., {Hinz}, P.~M., {et~al.} 2003{\natexlab{b}}, \aj,
  125, 1458

\bibitem[{Smith {et~al.}(2004)Smith, Vink, \& de~Koter}]{Smith04}
Smith, N., Vink, J.~S., \& de~Koter, A. 2004, \apj, 615, 475

\bibitem[{{St.-Louis} {et~al.}(1987){St.-Louis}, {Drissen}, {Moffat},
  {Bastien}, \& {Tapia}}]{StL87}
{St.-Louis}, N., {Drissen}, L., {Moffat}, A.~F.~J., {Bastien}, P., \& {Tapia},
  S. 1987, \apj, 322, 870

\bibitem[{Stahl(1987)}]{Stahl87}
Stahl, O. 1987, \aap, 182, 229

\bibitem[{Stahl {et~al.}(2003)Stahl, Gäng, Sterken, Kaufer, Rivinius,
  Szeifert, \& Wolf}]{Stahl03}
Stahl, O., Gäng, T., Sterken, C., {et~al.} 2003, \aap, 400, 279

\bibitem[{Stahl {et~al.}(2001)Stahl, Jankovics, Kovács, Wolf, Schmutz, Kaufer,
  Rivinius, \& Szeifert}]{Stahl01}
Stahl, O., Jankovics, I., Kovács, J., {et~al.} 2001, \aap, 375, 54

\bibitem[{{Stahl} {et~al.}(1983){Stahl}, {Wolf}, {Klare}, {Cassatella},
  {Krautter}, {Persi}, \& {Ferrari-Toniolo}}]{Stahl83}
{Stahl}, O., {Wolf}, B., {Klare}, G., {et~al.} 1983, \aap, 127, 49

\bibitem[{Sterken {et~al.}(1998)Sterken, de~Groot, \& van Genderen}]{Sterken98}
Sterken, C., de~Groot, M., \& van Genderen, A.~M. 1998, \aap, 333, 565

\bibitem[{Sterken {et~al.}(1991)Sterken, Gosset, Juttner, Stahl, Wolf, \&
  Axer}]{Sterken91}
Sterken, C., Gosset, E., Juttner, A., {et~al.} 1991, \aap, 247, 383

\bibitem[{Taylor {et~al.}(1991)Taylor, Nordsieck, Schulte-Ladbeck, \&
  Bjorkman}]{Taylor91}
Taylor, M., Nordsieck, K.~H., Schulte-Ladbeck, R.~E., \& Bjorkman, K.~S. 1991,
  \aj, 102, 1197

\bibitem[{van Boekel {et~al.}(2003)van Boekel, Kervella, Sch$\ddot{\rm o}$ller,
  Herbst, Brandner, de~Koter, Waters, Hillier, Paresce, Lenzen, \&
  Lagrange}]{vB03}
van Boekel, R., Kervella, P., Sch$\ddot{\rm o}$ller, M., {et~al.} 2003, \aap,
  410, 37

\bibitem[{van Genderen(2001)}]{vG01}
van Genderen, A.~M. 2001, A\&A, 366, 508

\bibitem[{van Genderen {et~al.}(1997)van Genderen, de~Groot, \& Sterken}]{vG97}
van Genderen, A.~M., de~Groot, M., \& Sterken, C. 1997, \aaps, 124, 517

\bibitem[{van Genderen \& Sterken(1999)}]{vG-S99}
van Genderen, A.~M. \& Sterken, C. 1999, \aap, 349, 537

\bibitem[{van Genderen {et~al.}(1992)van Genderen, Th\'{e}, de~Winter,
  Hollander, de~Jong, van~den Bosch, Kolkman, \& Verheijen}]{vG92}
van Genderen, A.~M., Th\'{e}, P.~S., de~Winter, D., {et~al.} 1992, \aap, 258,
  316

\bibitem[{{Vink} \& {de Koter}(2002)}]{V-dK02}
{Vink}, J.~S. \& {de Koter}, A. 2002, \aap, 393, 543

\bibitem[{Vink {et~al.}(1999)Vink, de~Koter, \& Lamers}]{Vink99}
Vink, J.~S., de~Koter, A., \& Lamers, H. J. G. L.~M. 1999, A\&A, 350, 181

\bibitem[{{Vink} {et~al.}(2002){Vink}, {Drew}, {Harries}, \&
  {Oudmaijer}}]{Vink02}
{Vink}, J.~S., {Drew}, J.~E., {Harries}, T.~J., \& {Oudmaijer}, R.~D. 2002,
  \mnras, 337, 356

\bibitem[{Vink {et~al.}(2003)Vink, Drew, Harries, Oudmaijer, \& Unruh}]{Vink03}
Vink, J.~S., Drew, J.~E., Harries, T.~J., Oudmaijer, R.~D., \& Unruh, Y.~C.
  2003, \aap, 403, 703

\bibitem[{Voors {et~al.}(1999)Voors, Waters, Morris, Trams, de~Koter, \&
  Bouwman}]{Voors99}
Voors, R. H.~M., Waters, L. B. F.~M., Morris, P.~W., {et~al.} 1999, \aap, 341,
  67

\bibitem[{{Waagen}(2005)}]{Waagen05}
{Waagen}, E.~O. 2005, private communication

\bibitem[{Weis(2003)}]{Weis03}
Weis, K. 2003, \aap, 408, 205

\bibitem[{Wolf {et~al.}(1974)Wolf, Campusano, \& Sterken}]{Wolf74}
Wolf, B., Campusano, L., \& Sterken, C. 1974, \aap, 36, 87

\bibitem[{{Wood} {et~al.}(1996){Wood}, {Bjorkman}, {Whitney}, \&
  {Code}}]{Wood96}
{Wood}, K., {Bjorkman}, J.~E., {Whitney}, B.~A., \& {Code}, A.~D. 1996, \apj,
  461, 828

\end{thebibliography}
\begin{onecolumn}
\appendix
\section{Atlas of early-type supergiants \label{sec:app}}
 \begin{figure}[h]
\centering
\includegraphics[bb=30 30 540 720,width=15cm,angle=0,clip]{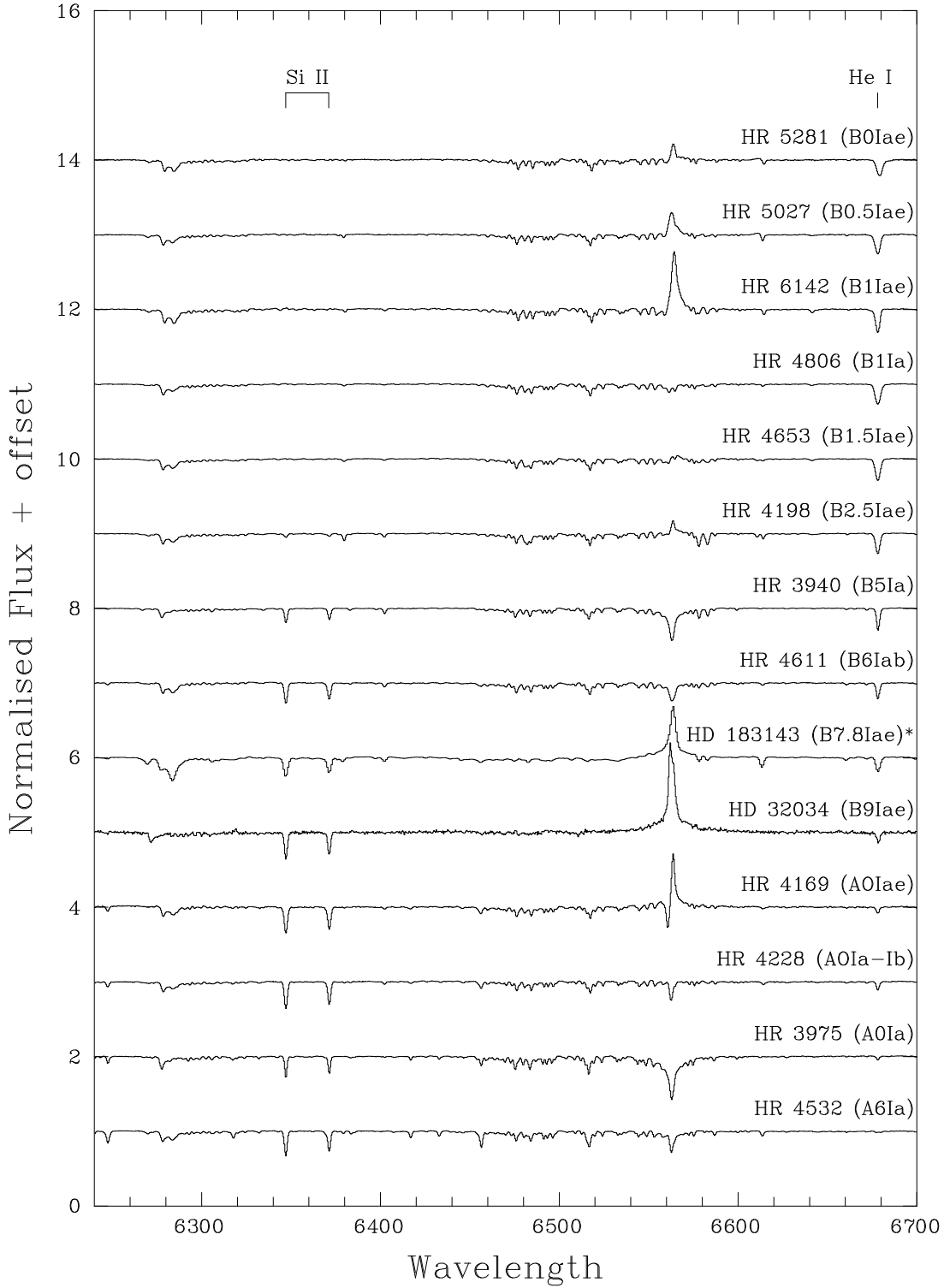}
 \caption[]{A spectral atlas of early-type supergiants in the range
 6240 - 6700 \AA. The prominent photospheric absorption lines of SiII
 and HeI, which appear to be good indicators of effective stellar
 temperature, have been identified. Spectral types of the stars were
 obtained from SIMBAD. The data in the atlas is available
 electronically through {\tt
 http://vizier.u-strasbg.fr/viz-bin/VizieR}.

*HD 183143 is a candidate LBV.

\label{fig:atlasspecs}}
\end{figure}
\end{onecolumn}

\end{document}